\documentclass[structabstract]{aa}  
\usepackage{graphicx}
\usepackage{txfonts}
\def\arcm{\hbox{$^\prime$}}  
\begin{document}
   \title{Statistical evaluation of the flux cross-calibration of the XMM-{\it Newton} EPIC cameras}

   \subtitle{}

   \author{
     S. Mateos  \inst{1}
     \and
     R.D. Saxton \inst{2}
     \and
     A.M. Read   \inst{1}
     \and
     S. Sembay   \inst{1}
}

   \offprints{sm279@star.le.ac.uk}

   \institute{
     Department of Physics and Astronomy, University of Leicester. University Road, Leicester, UK
     \and XMM SOC, ESAC, Apartado 78, 28691 Villanueva de la Ca\~{n}ada, Madrid, Spain
             }

   \date{26 January 2009}

 
  \abstract
  {The second XMM-{\it Newton} serendipitous source catalogue, {\tt 2XMM}, provides the ideal data base for performing a statistical 
    evaluation of the flux cross-calibration of the XMM-{\it Newton} European Photon Imaging Cameras (EPIC).
    }
  {We aim to evaluate the status of the relative flux calibration of the EPIC cameras on board the XMM-{\it Newton} observatory (MOS1, MOS2,
   and pn) and investigate the dependence of the calibration on energy (from 0.2 to 12 keV),
   position of the sources in the field of view of the X-ray detectors, and lifetime of the mission.
  }
  {We compiled the distribution of flux percentage differences for large samples of `good quality' objects detected with at
   least two of the EPIC cameras. The mean offset of the fluxes and dispersion of the distributions was then found by Gaussian fitting.
   Count rate to flux conversion was performed with a fixed spectral model. The impact on the results of varying this model was investigated.
    }
  {Excellent agreement was found between the two EPIC MOS cameras to better than 4\% over the entire energy
    range where the EPIC cameras are best calibrated (0.2-12.0 keV). 
    We found that MOS cameras register 7-9\% higher flux than EPIC pn below 4.5 keV and a 10-13\%
    flux excess at the highest energies ($\gtrsim$ 4.5 keV). No evolution of the flux ratios is seen with time, except at the lowest 
    energies ($\lesssim$ 0.5 keV), where we found a strong decrease in the MOS to pn flux ratio with time. 
    This effect is known to be due to a gradually degrading MOS redistribution function. The flux 
    ratios show some dependence on distance from the optical axis in the sense that the MOS to pn flux excess increases 
    with off-axis angle. Furthermore, in the 4.5-12.0 keV band there is a strong dependence of the MOS 
    to pn excess flux on the azimuthal-angle. These results strongly suggest that the calibration of the Reflection Grating Array (RGA) 
    blocking factors is incorrect at high energies. Finally, we recommend ways to improve the calculation of 
    fluxes in future versions of XMM-{\it Newton} serendipitous source catalogues.
    }
   {}


   \maketitle
%

\section{Introduction}
\label{intro}
ESA's XMM-{\it Newton} observatory (Jansen et al.~\cite{Jansen01}) was launched in December 1999. XMM-{\it Newton} 
carries on board three identical Wolter type 1 telescopes with 58 nested mirror shells. At the focal plane of 
each mirror there is a European Photon Imaging Camera (EPIC) that records images and spectra of celestial X-ray sources 
(Str\"{u}der et al.~\cite{Struder01}; Turner et al.~\cite{Turner01}). Two of the cameras use front-illuminated EPIC-MOS 
(Metal-Oxide Semi-conductor) CCDs as X-ray detectors (hereafter, MOS1 and MOS2), while the third camera 
uses an EPIC-pn (p-n-junction) CCD (hereafter, pn). EPIC has a field-of-view (FOV) of 30$\arcm$ diameter, moderate energy resolution and 
can be operated in various observational modes related to the readouts in each mode 
(Kendziorra et al.~\cite{Kendziorra97}; Kendziorra et al.~\cite{Kendziorra99}; Kuster et al.~\cite{Kuster99}; Ehle et al.~\cite{Ehle01}). The two telescopes with the MOS cameras at the focal plane have a Reflection Grating 
Spectrometer mounted behind the mirrors (RGS, den Herder et al.~\cite{Herder01}). The incident light is split 
roughly 50\%:50\% between the RGS and MOS cameras.

The absolute and relative calibration of the XMM-{\it Newton} instruments has been monitored 
and studied since launch. These studies are mostly based on detailed spectral analyses of bright `calibration'
targets that have been observed regularly by XMM-{\it Newton}. The relative flux 
calibration of the EPIC cameras is known to be better than 7\% on-axis, with MOS fluxes being higher than pn fluxes (see Stuhlinger et al.~\cite{Stuhlinger08}), and has been 
shown to have both time and energy dependence. 
For details on the calibration of 
other XMM-{\it Newton} science instruments see Stuhlinger et al.~(\cite{Stuhlinger08}). 

In this work we use a large sample of serendipitous X-ray sources to study 
the flux cross-calibration of the EPIC cameras as a function of 
various parameters. 
We investigated the performance of the EPIC instruments by carrying out a systematic analysis of the 
camera-to-camera flux ratio for a large number of sources with `good quality' X-ray data in various energy bands 
over the XMM-{\it Newton} lifetime. 
We compare the fluxes measured by the EPIC cameras in different energy bands covering the 
entire energy range from 0.2 keV to 12.0 keV, for sources detected at different epochs 
and detected at different positions in the field of view. In this way we can investigate 
how instrumental effects such as the detector quantum efficiency, point spread function (PSF) and mirror vignetting function 
affect the relative flux calibration of the EPIC cameras. 
This is an important complement to studies based upon bright targets, which are necessarily restricted to 
the central CCD of the MOS cameras. The outer MOS CCDs are operated exclusively in {\tt FullFrame} mode 
(i.e. {\tt PrimeFullWindow} mode) which has a pile-up threshold 
of 0.7 cts/s\footnote{See the on line XMM-{\it Newton} User Handbook 
{\tt http://xmm.esac.esa.int/external/xmm\_user\_support/
documentation/uhb/}}. 

For this analysis we have used the sources in the second XMM-{\it Newton} serendipitous 
source catalogue, {\tt 2XMM} (Watson et al.~\cite{Watson08}). {\tt 2XMM} is the largest catalogue of 
X-ray sources compiled to date, containing more than 240 000 sources, and therefore provides the ideal 
data base to conduct this study. Furthermore, we have used the same energy band definition as in 
{\tt 2XMM}: 0.2-0.5 keV, 0.5-1.0 keV, 1.0-2.0 keV, 2.0-4.5 keV and 4.5-12.0 keV.

This paper is organised as follows:
In \textsection2 we give a short description of the source 
detection procedure used to compile the {\tt 2XMM} catalogue (\textsection2.1) 
and present the criteria for selection of sources for this analysis 
(\textsection2.2). 
In \textsection3 we explain the approach used to calculate the fluxes of the objects.
The main results of this work are presented in \textsection4 and discussed in 
\textsection5. Finally, the conclusions are summarised in \textsection6.

\begin{table*}
  \caption{Count rate to flux energy conversion factors (ecf)$^a$.}
\label{table:1}      
\centering                          
\begin{tabular}{c c c c c c c c c c c c c c c c}        
\hline\hline                 
Energy band     &   & Open & & & Thin & & & Medium & & & Thick  \\
keV & ${\rm ecf_{pn}}$ & ${\rm ecf_{m1}}$ & ${\rm ecf_{m2}}$ & ${\rm ecf_{pn}}$ & ${\rm ecf_{m1}}$ & ${\rm ecf_{m2}}$ & ${\rm ecf_{pn}}$ & ${\rm ecf_{m1}}$ & ${\rm ecf_{m2}}$ & ${\rm ecf_{pn}}$ & ${\rm ecf_{m1}}$ & ${\rm ecf_{m2}}$ \\
\hline                        
    0.2-0.5   &   16.530   &    3.094   &    3.142   &    9.056   &    1.725   &    1.738   &    7.880   &    1.514   &    1.521   &    4.699   &    0.985   &    0.983        \vspace{0.05cm}\\
      0.5-1.0   &   10.256   &    2.191   &    2.196   &    8.253   &    1.807   &    1.810   &    7.995   &    1.756   &    1.759   &    6.138   &    1.426   &    1.428        \vspace{0.05cm}\\
        1.0-2.0   &    6.166   &    2.125   &    2.132   &    5.878   &    2.028   &    2.034   &    5.779   &    1.993   &    1.999   &    4.999   &    1.775   &    1.781        \vspace{0.05cm}\\
      2.0-4.5   &    1.982   &    0.758   &    0.762   &    1.950   &    0.747   &    0.751   &    1.927   &    0.738   &    0.742   &    1.827   &    0.708   &    0.712        \vspace{0.05cm}\\
     4.5-12.0   &    0.556   &    0.144   &    0.151   &    0.555   &    0.144   &    0.151   &    0.554   &    0.143   &    0.151   &    0.548   &    0.141   &    0.149        \vspace{0.05cm}\\
\hline                                   
\end{tabular}

$^a$ Energy conversion factors have been calculated for each camera, optical blocking filter of the observation and energy band.
To obtain the values a power-law spectral model of photon index $\Gamma$=1.7 and observed absorption 
${\rm N_H=3\times10^{20}\,cm^{-2}}$ was assumed. The {\tt ecf} are given in units of ${\rm 10^{11}\,cts\,cm^2\,erg^{-1}}$.
\end{table*}

\section{The XMM-{\it Newton} data}
\label{data}
In order to carry out our analysis we used the 3491 observations included in the 
second XMM-{\it Newton} serendipitous source catalogue, {\tt 2XMM}. 
The catalogue contains observations from XMM-{\it Newton} up to revolution 1338 
and includes all the main MOS and pn observing modes. Only observations that were publicly 
available by 2007 May 01 are included in the catalogue\footnote{XMM-{\it Newton} started taking scientific observations in January 2000.}. 
{\tt 2XMM} contains 246\,897 X-ray source detections (sources with an EPIC 0.2-12.0 keV 
detection likelihood $\ge$6) of which 191\,870 are unique X-ray sources. 
Extended sources comprise $\sim$8\% of the detections. 
The data was processed with the XMM-{\it Newton} Science Analysis 
Software ({\tt SAS v7.1.0}\footnote{{\tt http://xmm.esac.esa.int/sas/8.0.0/}}) and 
a constant pipeline configuration. 
This guarantees that we have a uniform data set. 

\subsection{Source detection}
\label{src_det}
The source detection algorithm used to make {\tt 2XMM}, {\tt eboxdetect}-{\tt emldetect},
is run on the three EPIC cameras and on five energy bands 
simultaneously: 0.2-0.5 keV, 0.5-1.0 keV, 1.0-2.0 keV, 2.0-4.5 keV 
and 4.5-12.0 keV\footnote{In the 4.5-12.0 keV band pn photons with energies 
between 7.8-8.2 keV were excluded in order to avoid the instrumental background 
produced by Cu K-lines (Lumb et al.~\cite{Lumb02}).}. 
Images, cleaned for periods of high background during the observations, are 
created for each camera and energy band with the {\tt SAS} task {\tt evselect}. 
For the MOS cameras events with PATTERN$\le$12 
(single and double events) are used at all energies while for pn, 
events with PATTERN=0 (single events) below 0.5 keV and 
with PATTERN$\le$4 (single and double events) above 0.5 keV are used instead.
Detection masks defining the area of the detector suitable for source detection   
are created with the {\tt SAS} task {\tt emask} for each camera.
Energy-dependent exposure maps are computed with the {\tt SAS} task {\tt eexpmap}.
The {\tt eexpmap} task uses the latest calibration information for the mirror vignetting, 
detector quantum efficiency and filter transmission to compute the telescope and 
instrumental throughput efficiency as a function of energy and position in the FOV. Quantum 
efficiency, filter transmission and vignetting are evaluated at the 
mean energy of each energy band.

Initial source lists are obtained using the {\tt SAS} task {\tt eboxdetect} 
by employing a simple sliding box cell detection algorithm. 
Smoothed background maps for each camera and energy band are 
produced by the {\tt SAS} task {\tt esplinemap} which performs a 
spline fit on the source-free images, after masking out all 
sources detected by {\tt eboxdetect}.
{\tt Eboxdetect} is run for a second time using the background maps 
produced by {\tt esplinemap}, which increases the sensitivity of 
the source detection. 

The final source lists and source parameter estimation are obtained 
with the {\tt SAS} task {\tt emldetect}. {\tt Emldetect} performs  
a maximum likelihood fitting of the distribution of 
counts of sources detected by {\tt eboxdetect} with the instrumental 
point spread function (PSF) over a circular area of 60$\arcsec$ radius 
centred at the source positions. The fitting is performed using a tabulated energy and position dependent PSF.
In addition, {\tt emldetect} carries out a fit with the PSF convolved with a 
beta-model profile in order to search for sources extended in X-rays. 
The free parameters of the fits are the source position, 
count rates and extent. Position and extent are constrained to the best-fit value 
for all cameras and energy bands, while count rates are fitted separately for each camera 
and energy band. {\tt Emldetect} uses the exposure maps to correct the count rates for 
instrumental effects (mirror vignetting, detector quantum efficiency and filter 
transmission). Count rates and therefore fluxes given by {\tt emldetect} are 
background subtracted and corrected for PSF losses, i.e. 
they correspond to the flux in the entire PSF.

   \begin{figure}
   \centering
   \includegraphics[angle=-90,width=0.45\textwidth]{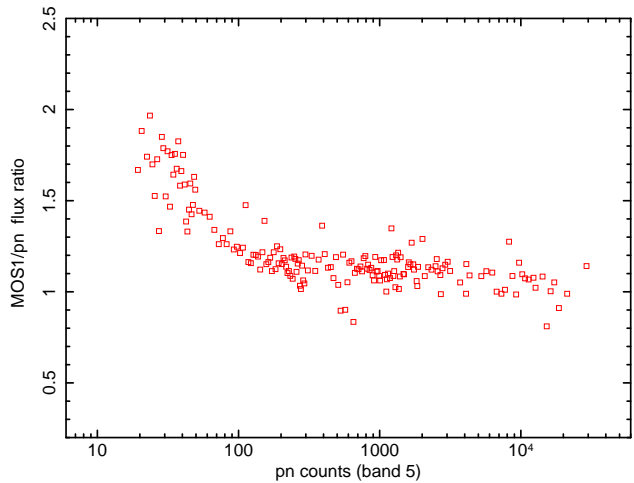}
   \caption{MOS1 to pn flux ratio as a function of 
     pn number of counts (background subtracted) in the 4.5-12.0 keV energy band.
     }
              \label{flux_rat}%
    \end{figure}

\subsection{Selection of sources}
The measured flux of a source can be significantly different between cameras for reasons other than a cross-calibration 
effect. For example a large discrepancy in the EPIC fluxes can be found if the source position falls in a CCD gap in one of the cameras. 
In order to reduce the number of `problematic' cases included in our analysis, we performed the following 
filtering to the source lists:
\begin{enumerate}
  \item We used only sources detected as point-like by the source detection algorithm.
  \item Sources with $<$200 background subtracted counts in the energy band of interest 
    and on each camera were excluded. 
    This requirement is used to avoid biases introduced by
    imposing a minimum value of the
    detection likelihood and insisting that the source is detected in both
    cameras (see Fig.~\ref{flux_rat}).
  \item Sources with off-axis angle\footnote{The off-axis angle is the separation of the 
    source position in the FOV from the optical axis.} 
    greater than 12$\arcm$ are excluded to ensure that 
    no azimuthal biases are introduced; beyond this radius, sources begin to fall outside of the MOS CCD boundaries at various different 
    azimuthal angles. 
    Sources within this radius that fall on CCD gaps are also excluded if 
      the detector coverage (weighted with the local PSF)
      is less than 15\% (these sources are considered non-detections in the {\tt 2XMM} 
      catalogue)\footnote{We have checked that the use of sources with small detector 
	coverage ($\lesssim$50\%) has no impact on the results of our analysis.}.
  \item Sources with a 2.0-12.0 keV observed flux ${\rm \ge6\times 10^{-12}\,erg\,cm^{-2}\,s^{-1}}$ have been excluded 
    from the analysis as these objects suffer from pile-up\footnote{If a source is bright 
enough, two or more X-ray photons 
      might deposit their charge in a single pixel or in neighbouring pixels during one read-out cycle. This is known as pile-up. 
      In such a case these events are recognised as one single event 
      with an energy equivalent to the sum of the contributing photon energies.
      Pile-up reduces the counts in the central part of the source, resulting in flux loss.} 
    and therefore their measured flux is underestimated.
\end{enumerate}

\section{Count rate to flux conversion}
\label{ecf}
The fluxes of the sources have been computed as  
\[
{\rm Flux={{\tt RATE} \over {\tt ecf}}}
\]
where {\tt RATE} are the source count rates in units of 
${\rm cts\,s^{-1}}$ and {\tt ecf} are the energy conversion factors from count rates to fluxes, in units 
of ${\rm cts\,cm^2\,erg^{-1}}$. Count rates are background subtracted and are corrected for PSF losses 
and variations of the effective exposure across the EPIC FOV. 

The energy conversion factors are different for each EPIC camera and depend on the optical blocking filter 
used for the observation and the observing mode\footnote{Each EPIC camera is equipped with a set of three separate 
filters, named thick, medium and thin. For a detailed description of the science modes of the EPIC cameras see 
the on line XMM-{\it Newton} User Handbook {\tt http://xmm.esac.esa.int/external/xmm\_user\_support/
documentation/uhb/}}. 
The dependence of the energy conversion 
factors on the observing mode is small (1.0-2.0\%), hence the {\tt ecf} values have only been computed for 
the {\tt PrimeFullWindow} mode, 
which is the observing mode most frequently used for XMM-{\it Newton} observations. 
The {\tt ecf} values have been computed for each EPIC camera, energy band and optical blocking filter.

   \begin{figure}
   \centering
   \hspace{2.9cm}\includegraphics[angle=90,width=0.48\textwidth]{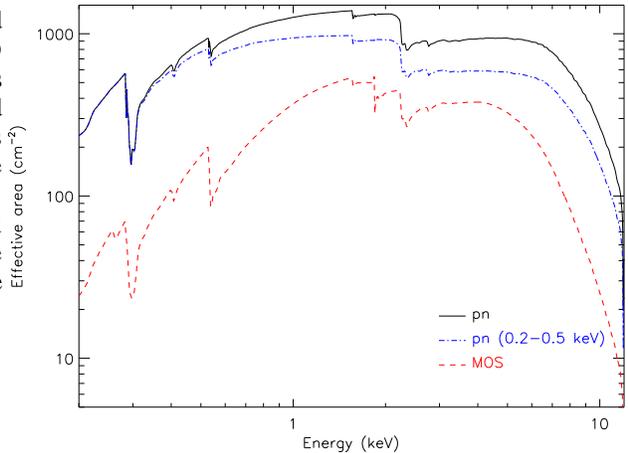}
   \caption{Variation of the effective area of the EPIC cameras (for the medium filter) as a function of energy. 
     The dot-dashed and solid lines show 
     the effective area for the EPIC pn camera, for single pixel 
     events (used below 0.5 keV) and for single+double pixel 
     events (used above 0.5 keV) respectively. The dashed line shows 
     the effective area for the EPIC MOS cameras.
     }
              \label{epic_arf}%
    \end{figure}

   \begin{figure*}
   \centering
   \hbox{
   \includegraphics[angle=90,width=0.48\textwidth]{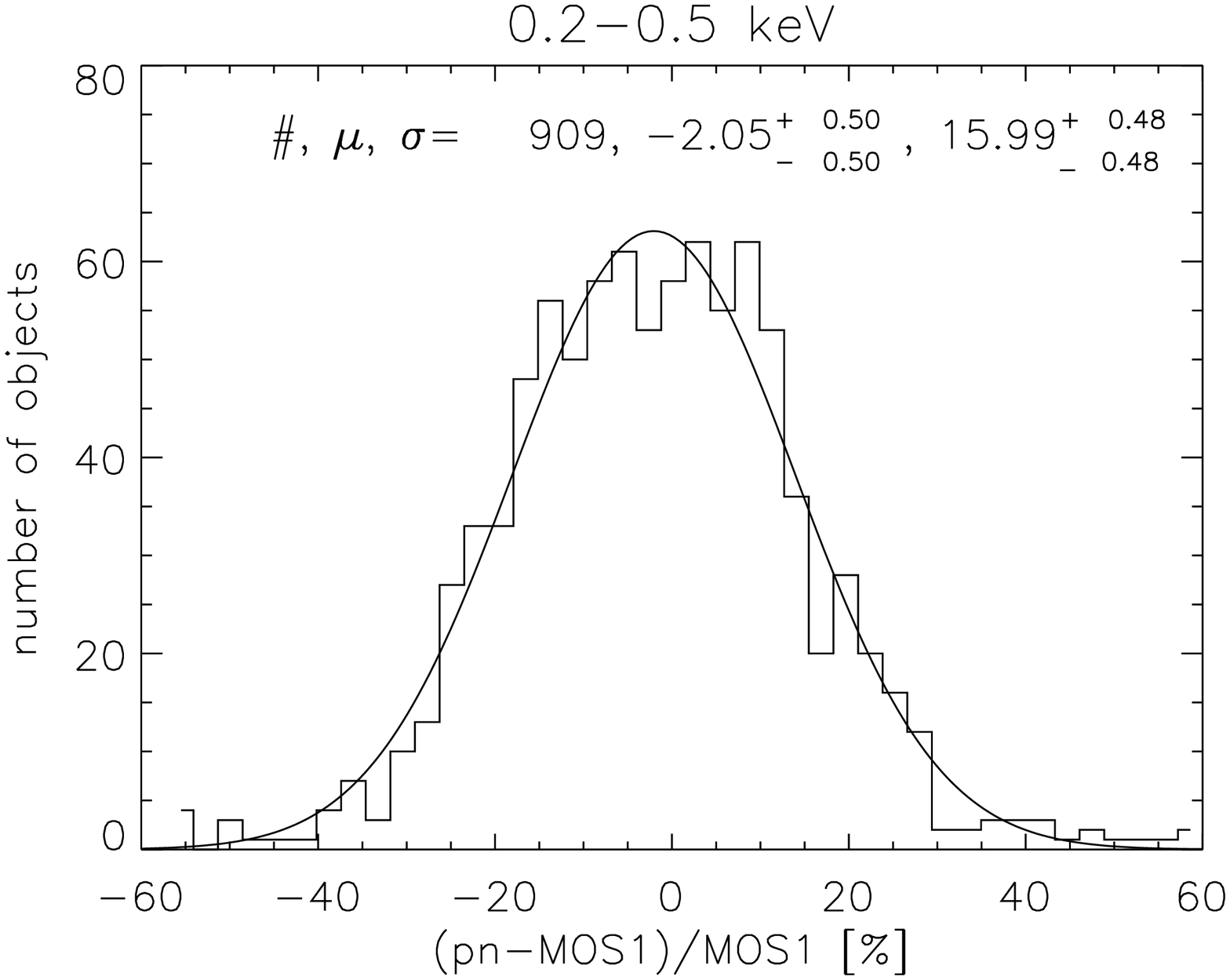}
   \includegraphics[angle=90,width=0.48\textwidth]{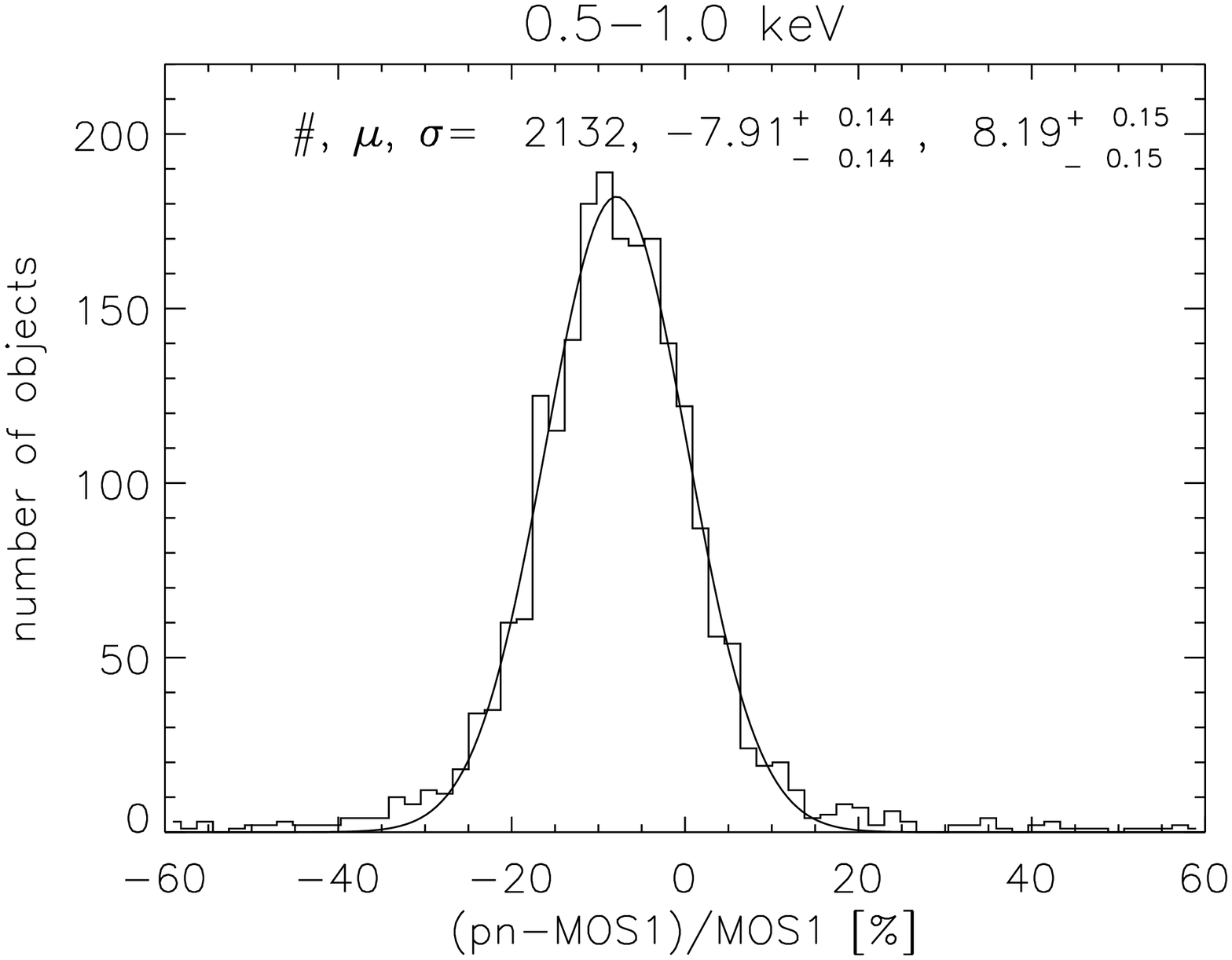}}
   \hbox{
   \includegraphics[angle=90,width=0.48\textwidth]{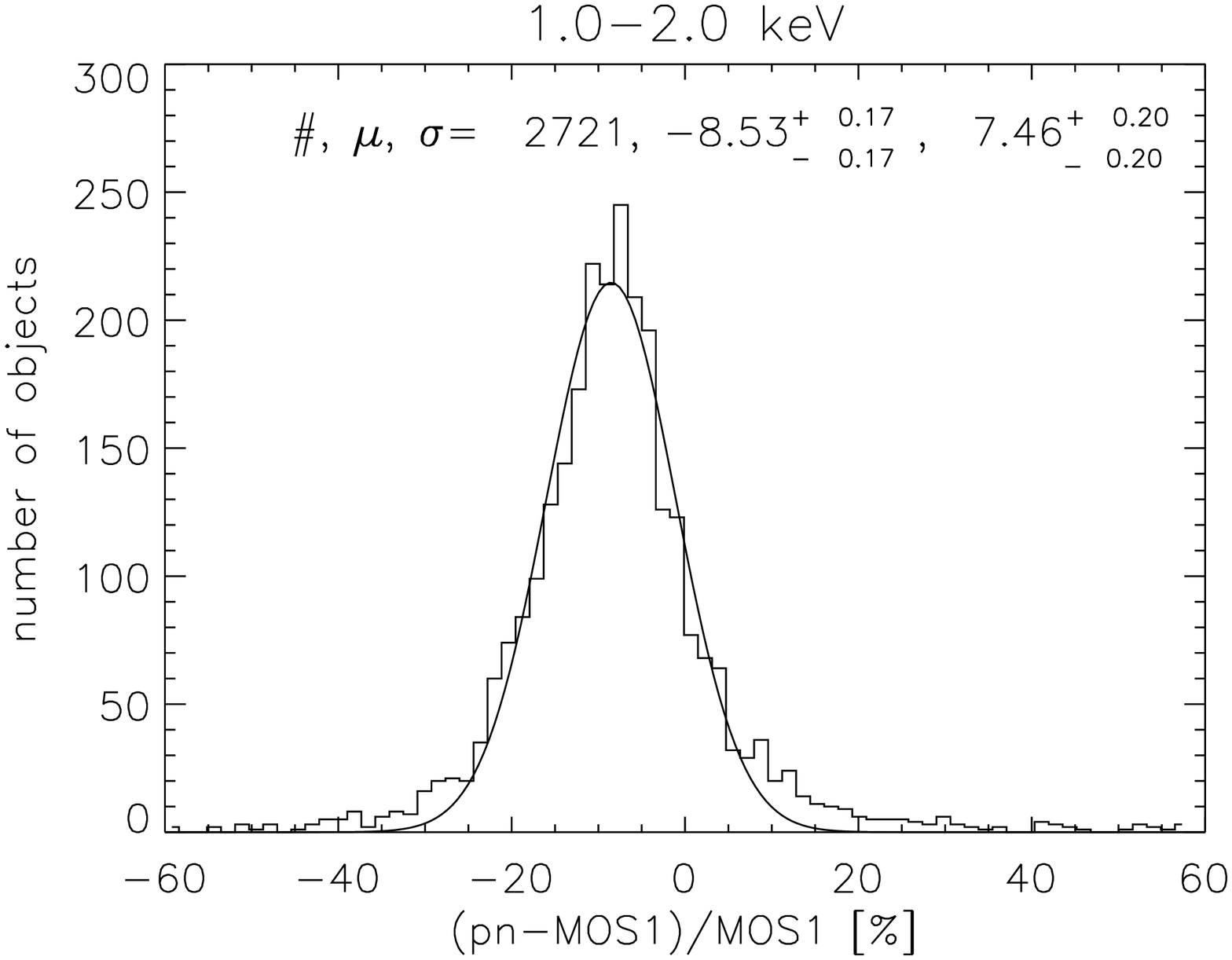}
   \includegraphics[angle=90,width=0.48\textwidth]{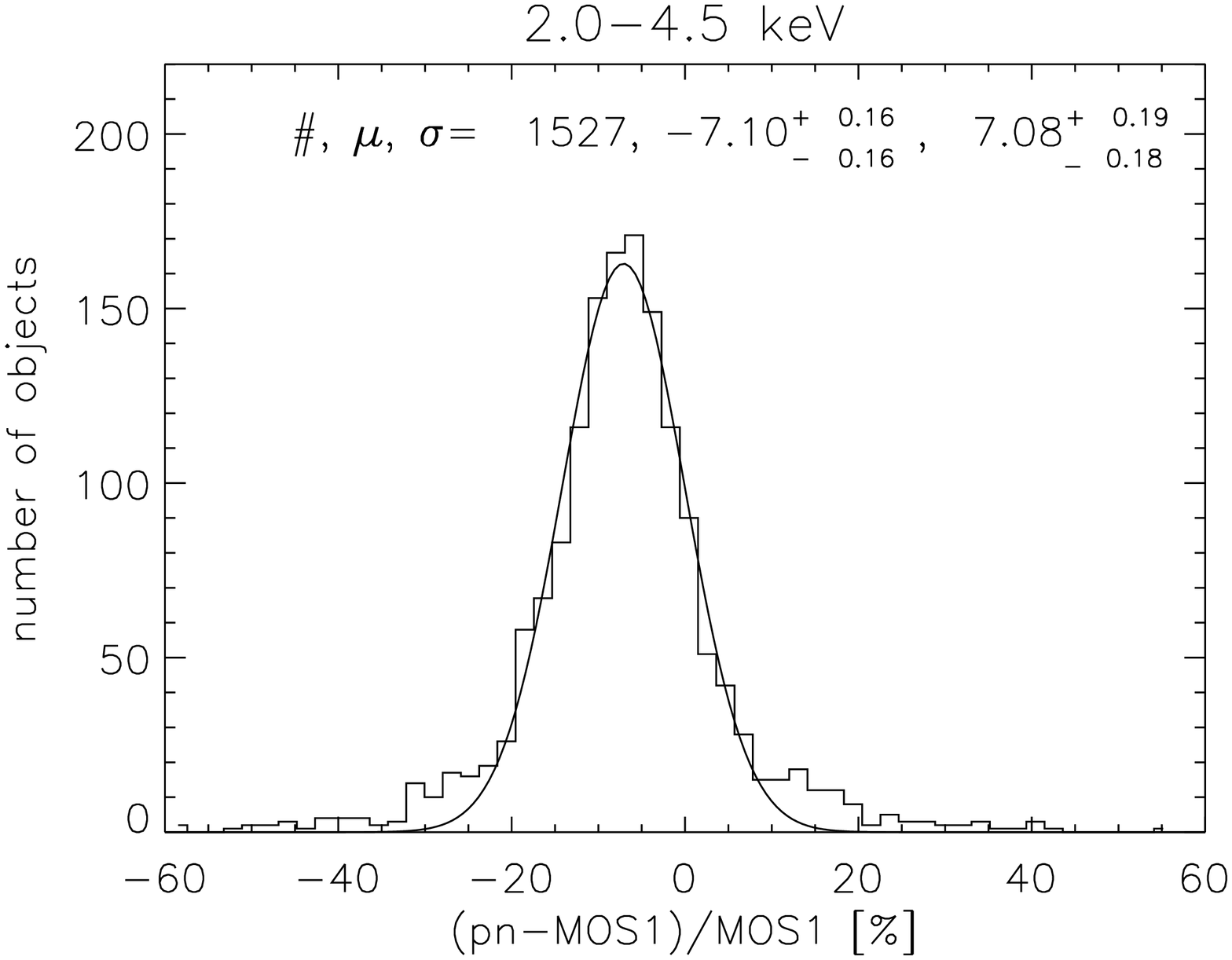}}
   \hbox{
   \includegraphics[angle=90,width=0.48\textwidth]{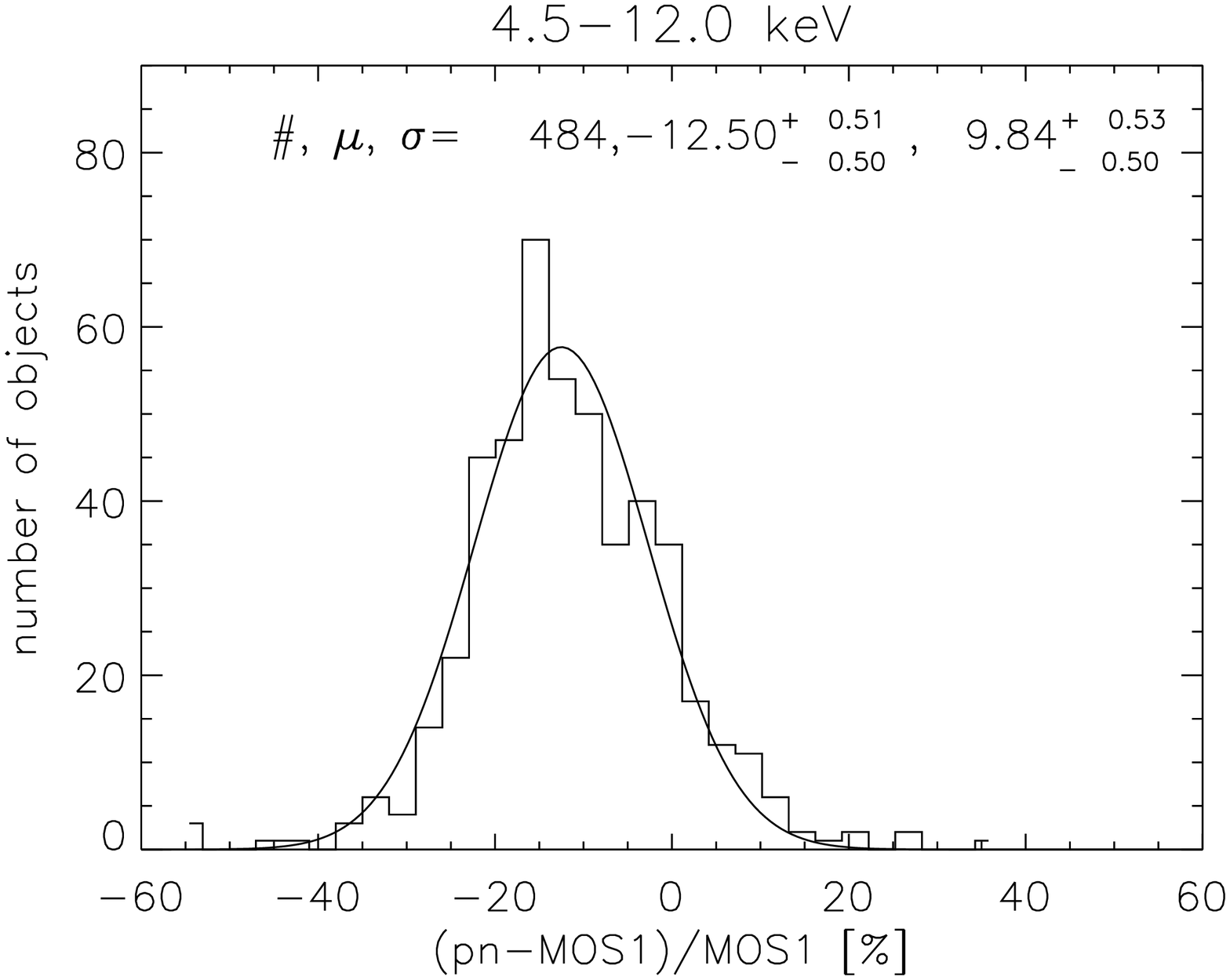}}
   \caption{Distributions of the percentage flux difference between the EPIC pn and MOS1 cameras as a function of the 
     energy band. The best-fit of the experimental distributions with a Gaussian function is shown with a smooth solid line and the corresponding best-fit 
     parameters (mean and dispersion) and number of sources involved in the analysis are listed in the upper-left corner of the plots 
     (also given in Table~\ref{table:2}). Errors are 90\% confidence.
     }
              \label{dists_pn_m1_all}%
    \end{figure*}

   \begin{figure*}
   \centering
   \hbox{
   \includegraphics[angle=90,width=0.48\textwidth]{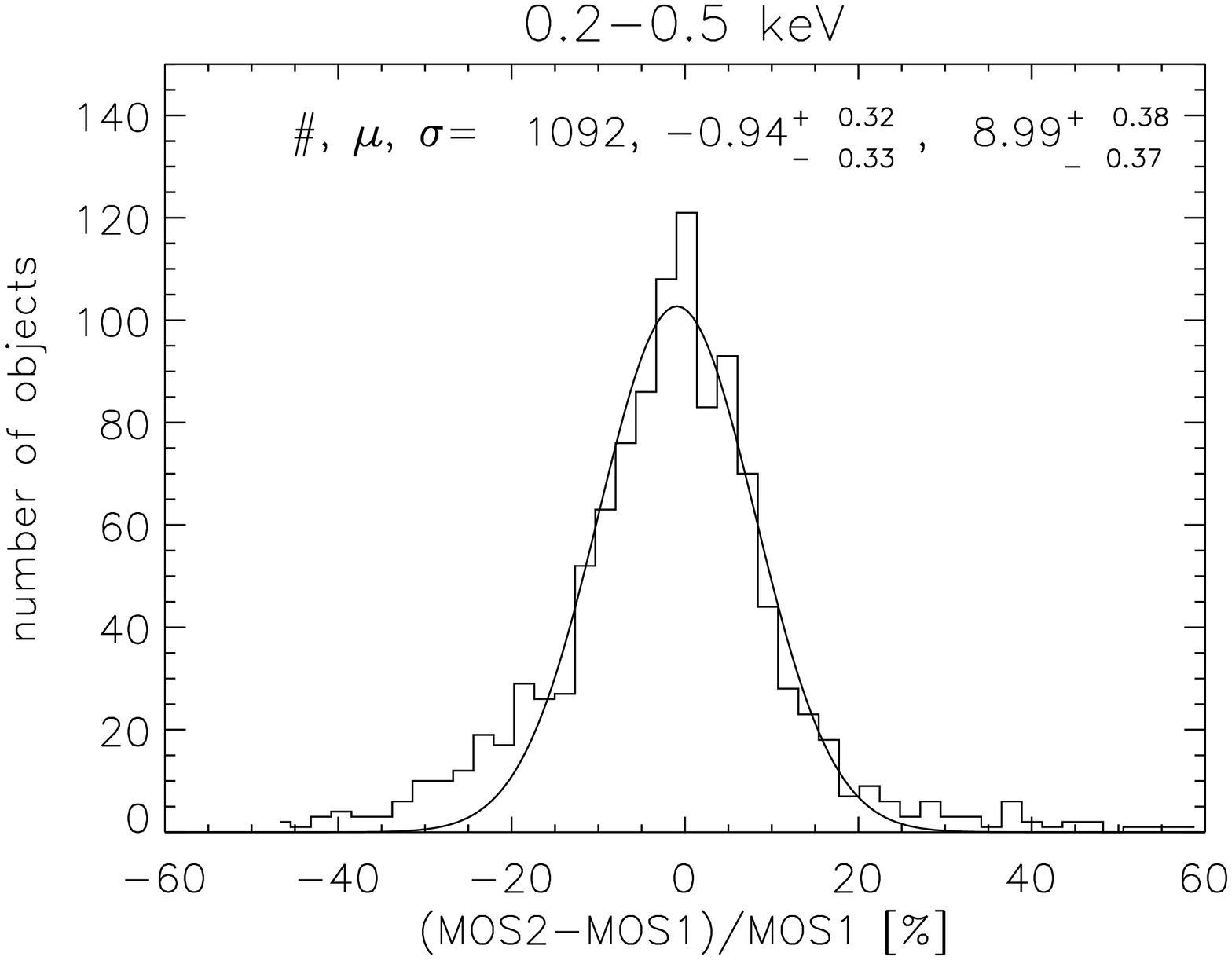}
   \includegraphics[angle=90,width=0.48\textwidth]{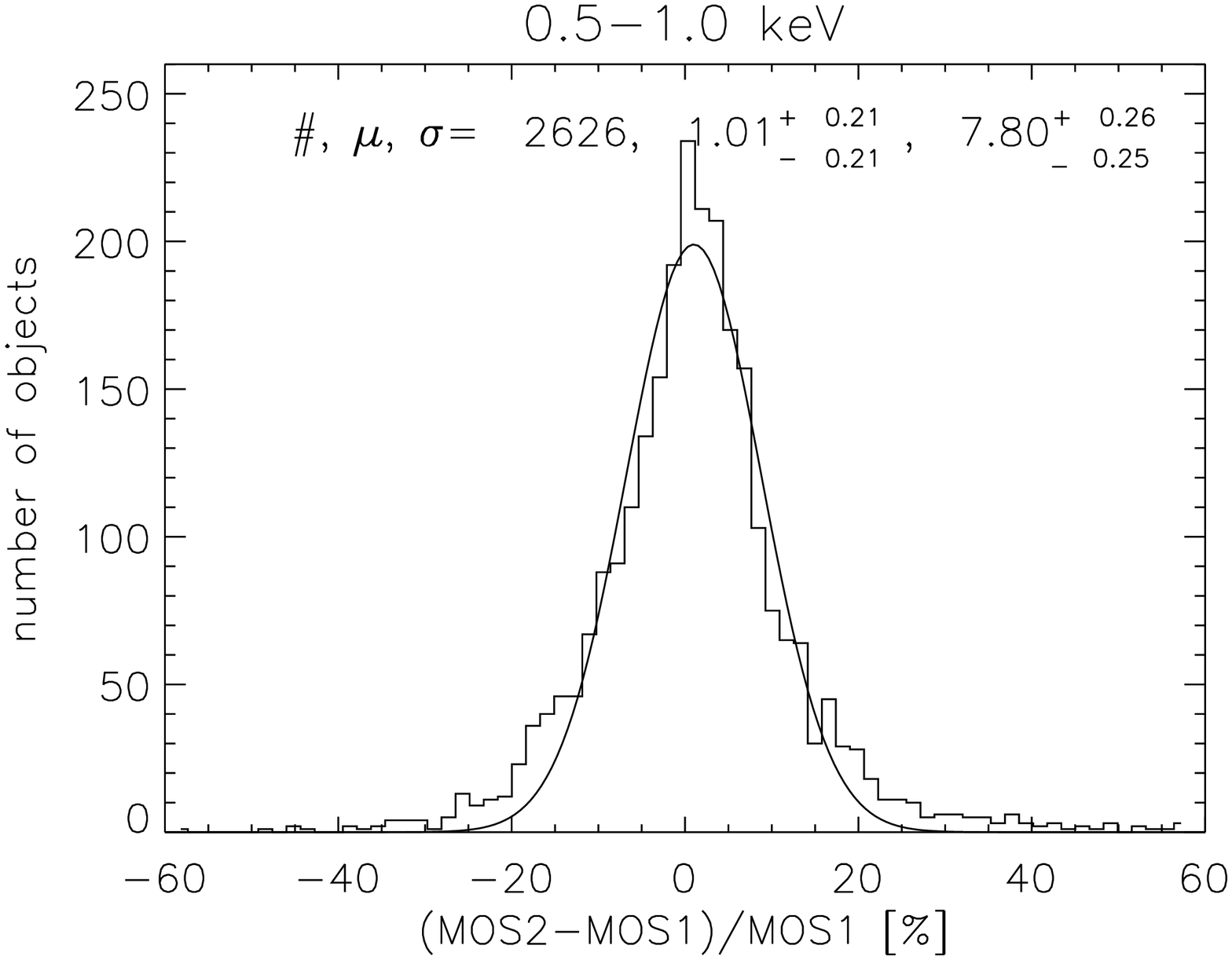}}
   \hbox{
   \includegraphics[angle=90,width=0.48\textwidth]{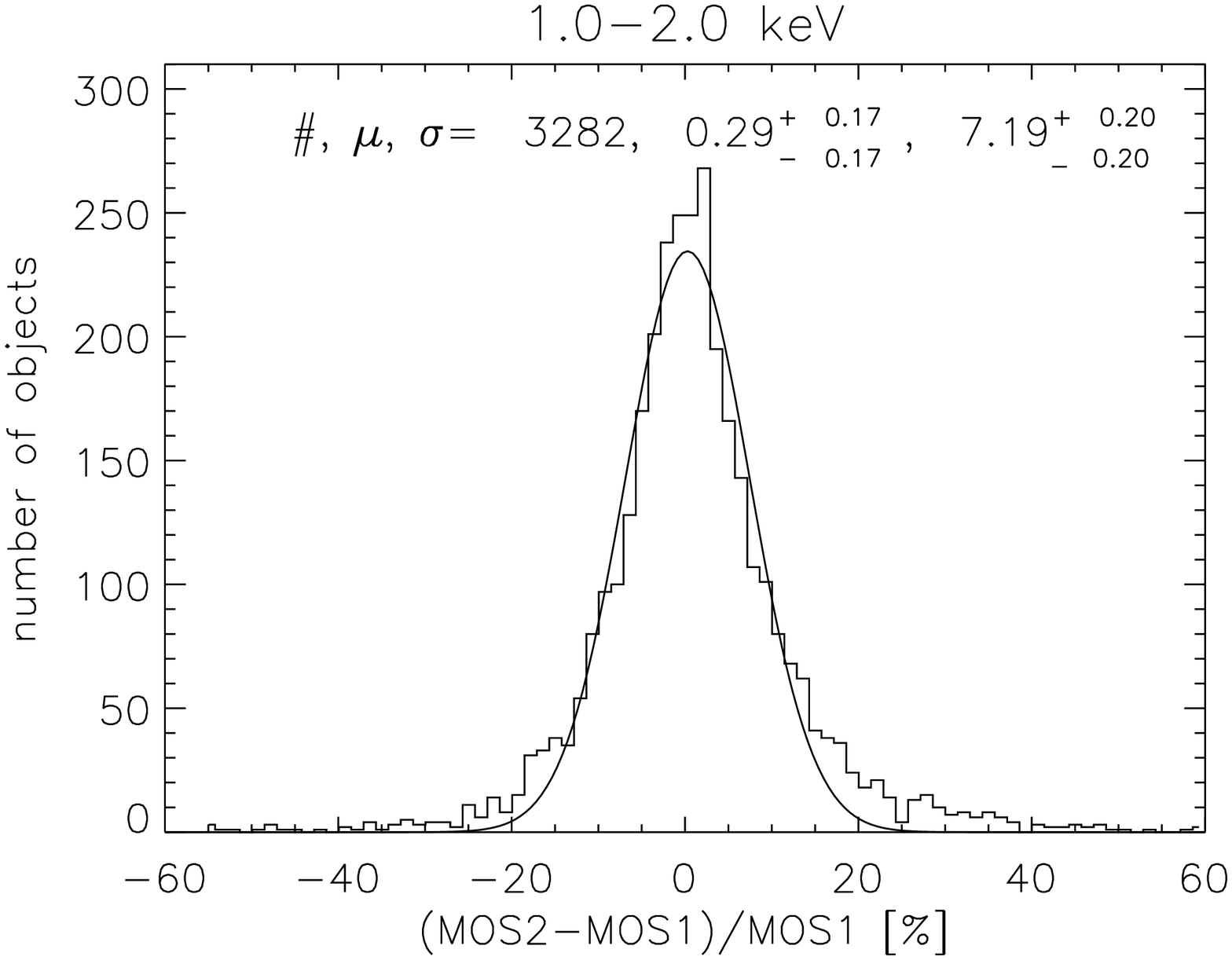}
   \includegraphics[angle=90,width=0.48\textwidth]{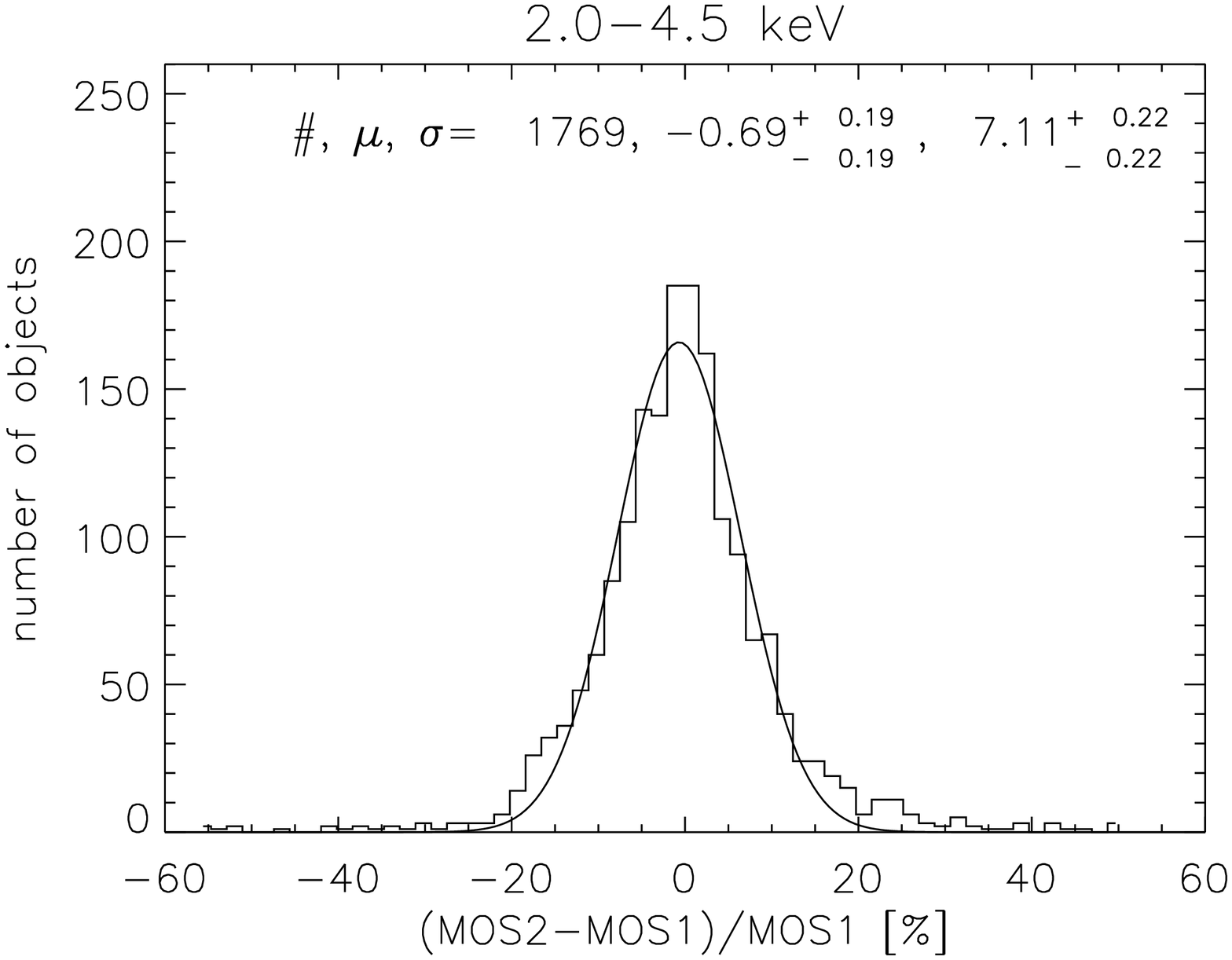}}
   \hbox{
   \includegraphics[angle=90,width=0.48\textwidth]{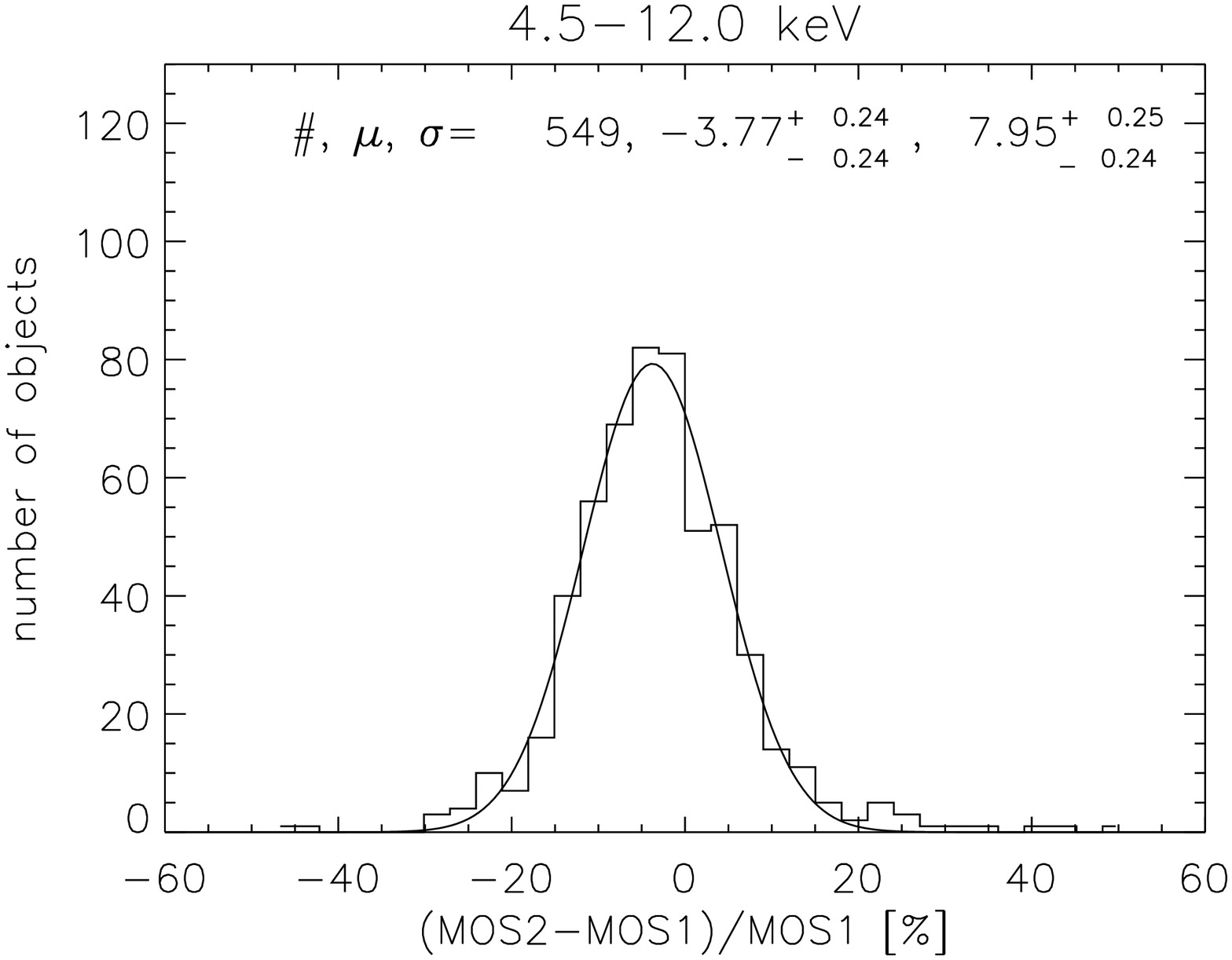}}
   \caption{Distributions of the percentage flux difference between the EPIC MOS1 and MOS2 cameras as a function of the 
     energy band. The best-fit of the experimental distributions with a Gaussian function is shown with a smooth solid line and the corresponding best-fit 
     parameters (mean and dispersion) and number of sources involved in the analysis are listed in the upper-left corner of the plots 
     (also given in Table~\ref{table:2}). Errors are 90\% confidence.
     }
              \label{dists_m2_m1_all}%
    \end{figure*}

   \begin{figure}[!h]
   \centering
   \hbox{
   \includegraphics[angle=90,width=0.5\textwidth]{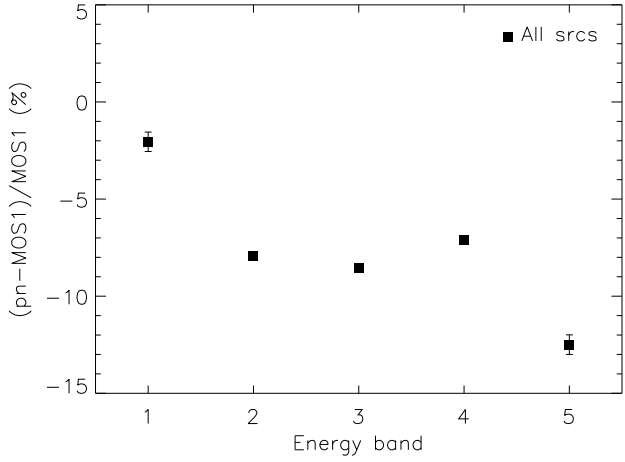}}
   \hbox{
   \includegraphics[angle=90,width=0.5\textwidth]{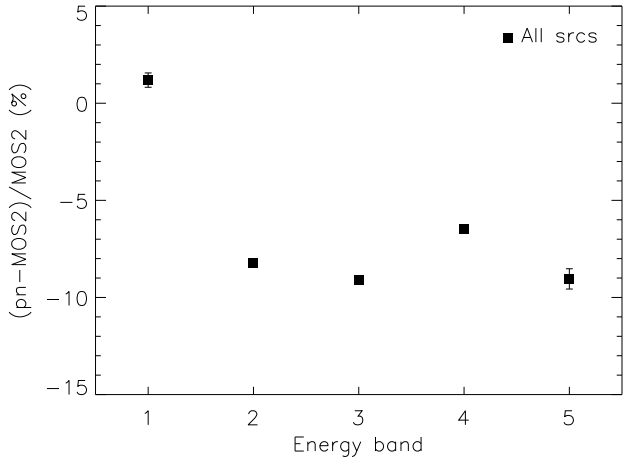}}
   \hbox{
   \includegraphics[angle=90,width=0.5\textwidth]{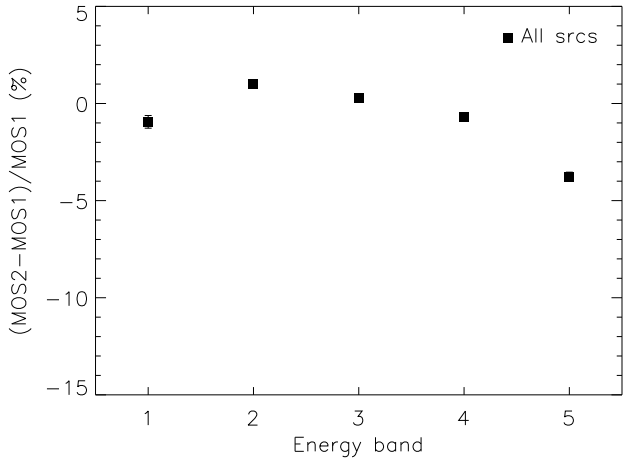}}
   \caption{Mean percentage flux difference between the EPIC cameras as a function of energy. 
     The x-axis shows the energy band identification number used in this 
     work: 1: 0.2-0.5 keV, 2: 0.5-1.0 keV, 3: 1.0-2.0 keV, 4: 2.0-4.5 keV and 5: 4.5-12.0 keV. 
     Top: pn-MOS1 comparison. Middle: pn-MOS2 comparison. Bottom: MOS2-MOS1 comparison. 
     Errors are 90\% confidence.
     }
              \label{disp_all_new_cal}%
    \end{figure}

   \begin{figure}[!ht]
   \centering
   \hbox{
   \includegraphics[angle=90,width=0.5\textwidth]{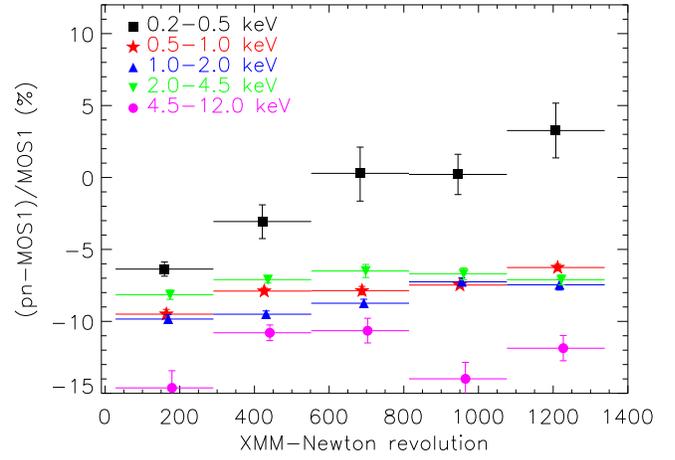}}
   \hbox{
   \includegraphics[angle=90,width=0.5\textwidth]{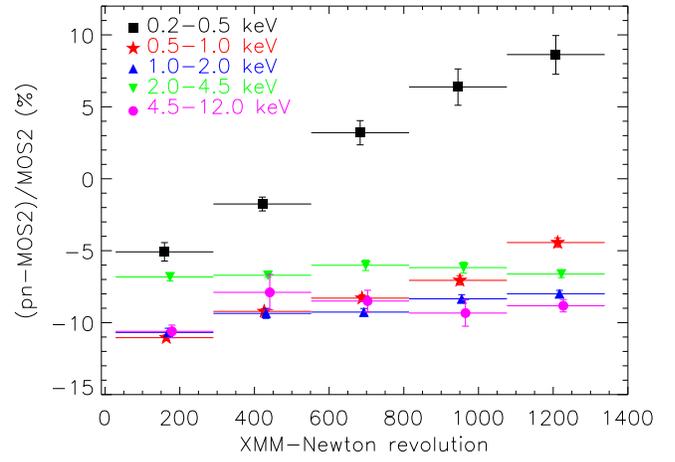}}
   \hbox{
   \includegraphics[angle=90,width=0.5\textwidth]{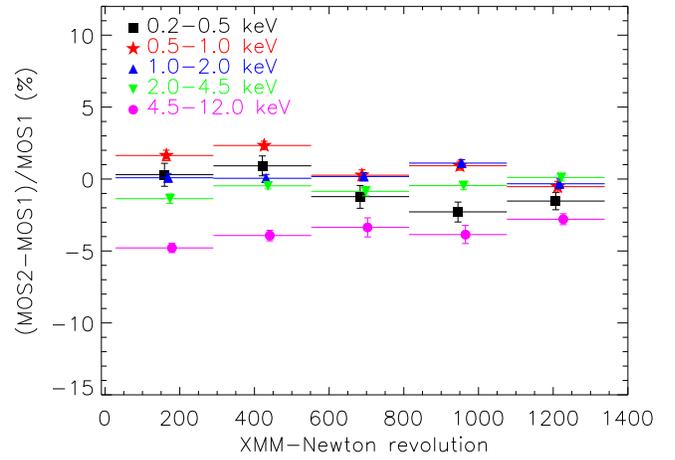}}
   \caption{Percentage flux difference between the EPIC cameras at different energies 
     as a function of time. Top: pn-MOS1 comparison.
     Middle: pn-MOS2 comparison. Bottom: MOS2-MOS1 comparison. Errors are 90\% confidence.
     }
   \label{var_flux_vs_rev}%
    \end{figure}

   \begin{figure}[!th]
   \centering
   \hbox{
   \includegraphics[angle=90,width=0.5\textwidth]{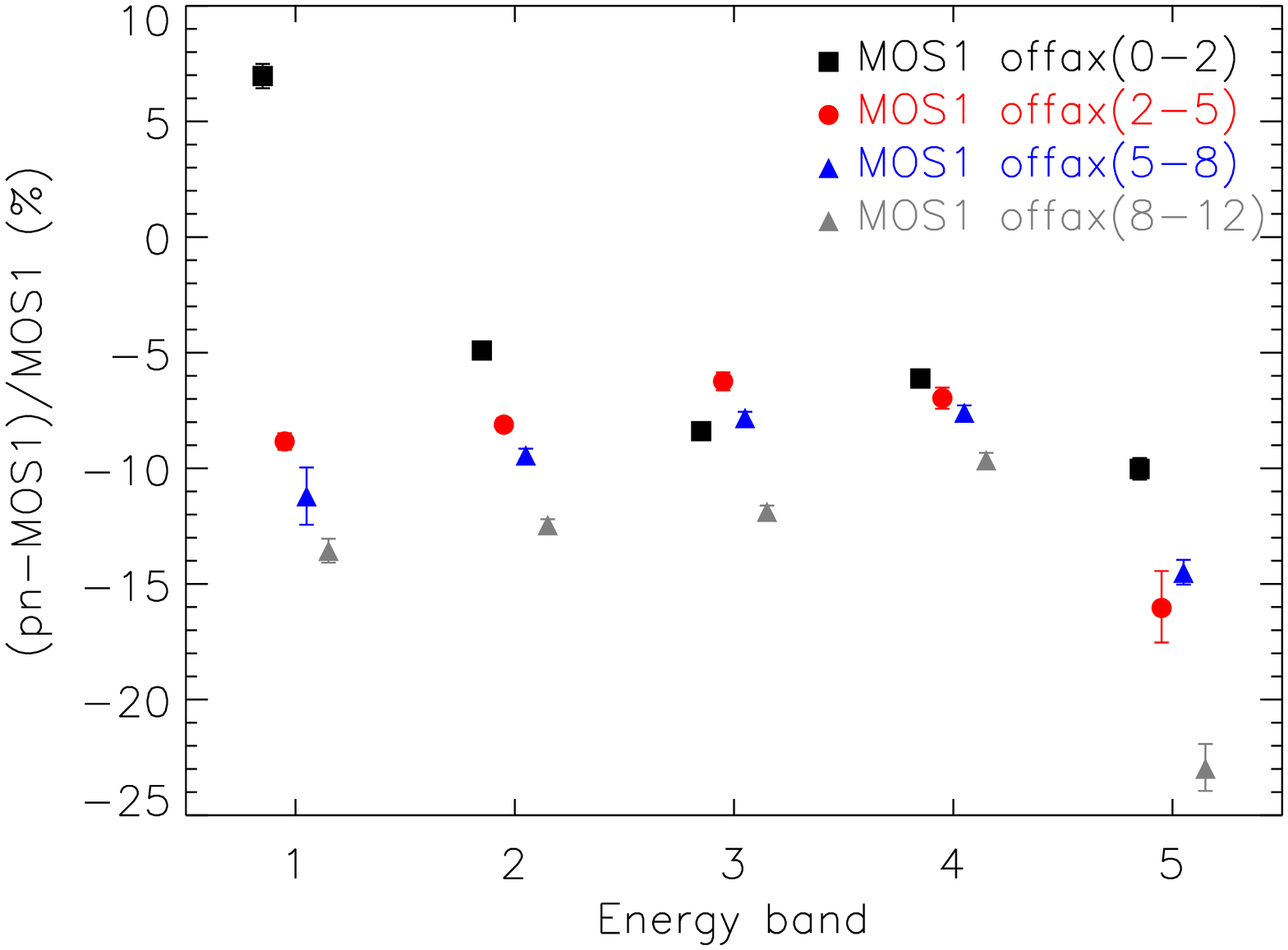}}
   \hbox{
   \includegraphics[angle=90,width=0.5\textwidth]{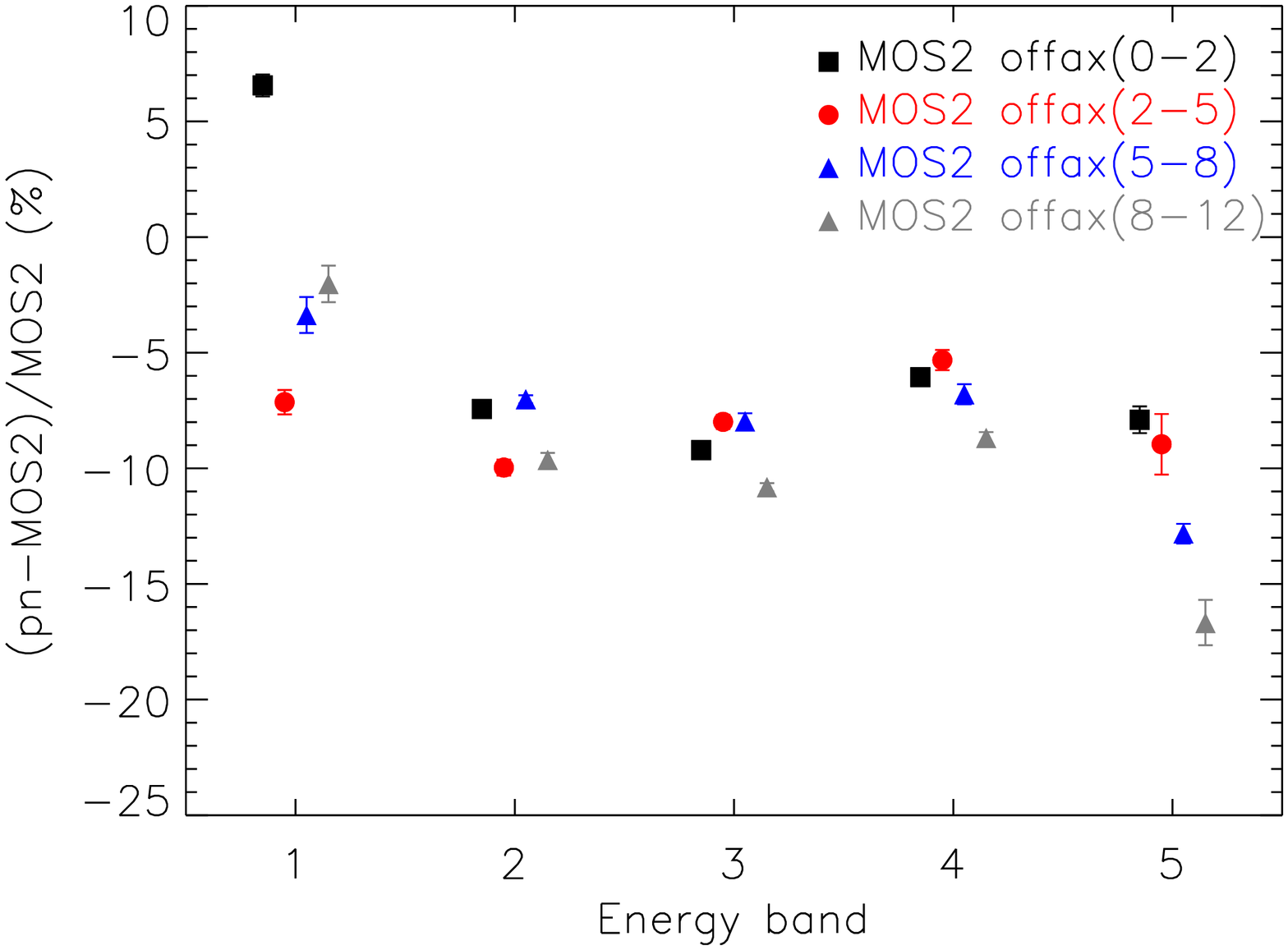}}
   \hbox{
   \includegraphics[angle=90,width=0.5\textwidth]{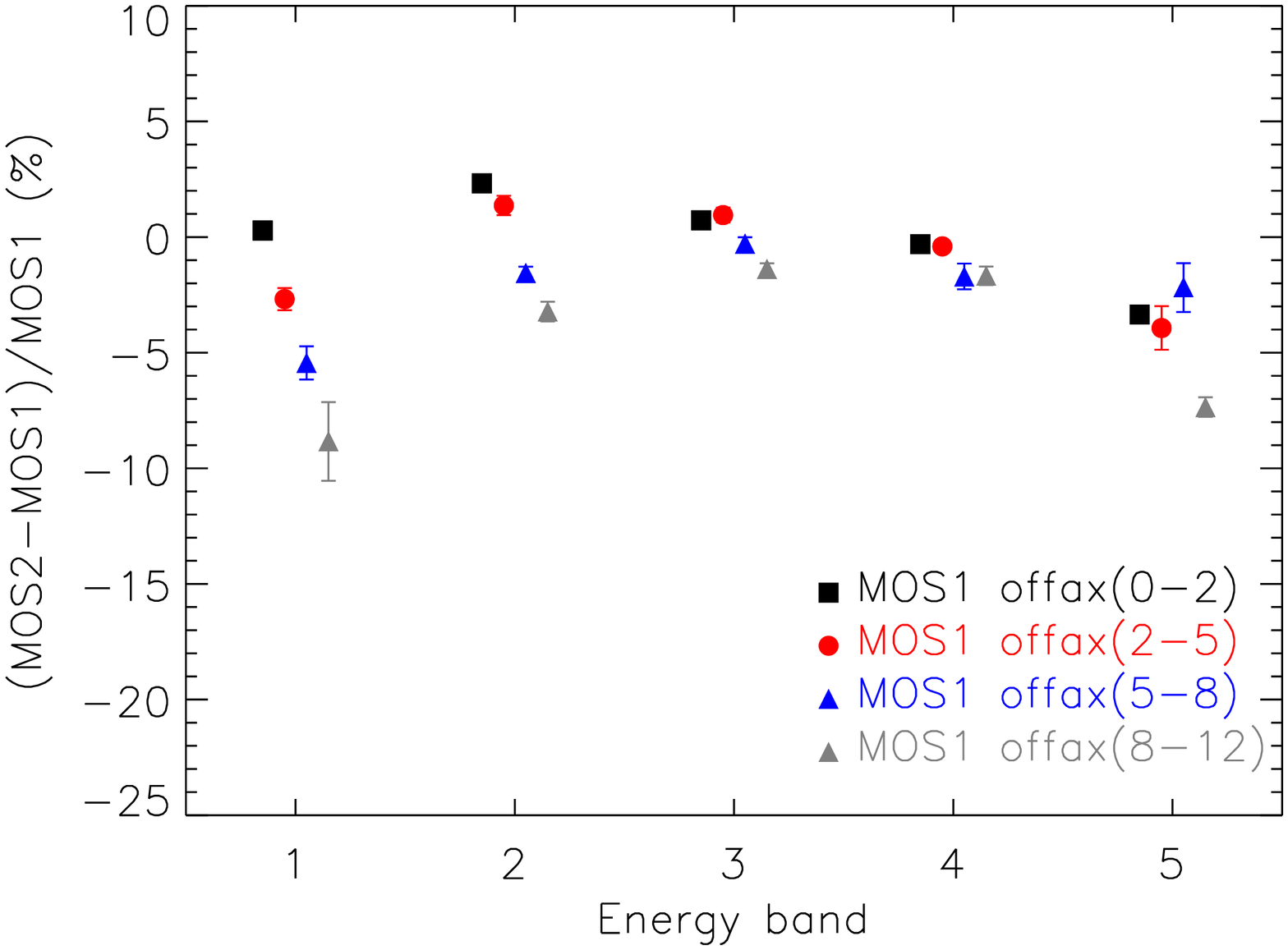}}
   \caption{Variation of the percentage flux difference between the EPIC cameras as a function of the 
     energy band for sources at different MOS off-axis angles.
     The x-axis shows the energy band identification number used in this 
     work: 1: 0.2-0.5 keV, 2: 0.5-1.0 keV, 3: 1.0-2.0 keV, 4: 2.0-4.5 keV and 5: 4.5-12.0 keV. 
     Data points at each energy band have been shifted slightly along the x-axis for clarity.
     Top: pn-MOS1 comparison. Middle: pn-MOS2 comparison. Bottom: MOS2-MOS1 comparison. 
     Errors are 90\% confidence.
     }
              \label{var_flux_vs_offaxis}%
    \end{figure}

In order to calculate the energy conversion factors it is necessary to assume a spectral model. 
In the case of {\tt 2XMM} and also for this work we have assumed that the model that best reproduces the broad 
band X-ray emission of the sources is a power-law with photon index $\Gamma=1.7$ and observed absorption 
${\rm N_H=3\times10^{20}\,cm^{-2}}$. This model has been found to be a good representation of the spectra of 
the bulk of {\tt 2XMM} sources (Watson et al.~\cite{Watson08}). Note that the fluxes have not been corrected for the Galactic absorption 
along the line of sight, and therefore they are absorbed fluxes.

Since the release of {\tt 2XMM} there have been important updates on the calibration information of the EPIC cameras.
For example, for the pn, new versions of the public redistribution matrix 
files ({\tt rmf}) are available. 
The most recent public calibration files\footnote{{\tt http://xmm2.esac.esa.int/external/xmm\_sw\_cal/calib/
epic\_files.shtml}} available at the time of writing, were used to create matrices
used in the computation of the energy conversion factors. These differ from the EPIC 
calibration files used in the pipeline processing of the {\tt 2XMM} catalogue. Note 
that correction factors to apply to the {\tt 2XMM} fluxes are provided in the on-line 
documentation of the catalogue\footnote{{\tt http://xmmssc-www.star.le.ac.uk/Catalogue/
xcat\_public\_2XMM.html}}. 
As indicated in Sec.~\ref{src_det}, the event selection used in the {\tt 2XMM} pipeline is for the pn detector: PATTERN$\le$4 at energies 
above 0.5 keV and PATTERN=0 only at energies below 0.5 keV. For the MOS cameras PATTERN$\le$12 selection is made at all 
energies. Therefore, for the EPIC pn detector 
we used the {\tt PrimeFullWindow} mode on-axis redistribution matrices for singles only 
(for the energy band 0.2-0.5 keV) and singles plus doubles (for the energy bands above 0.5 keV). 
The fluxes given by the EPIC pn camera from the new calibration are 
overall $\sim$2\% lower than {\tt 2XMM} fluxes at energies below 0.5 keV while at higher energies the EPIC pn flux changes are $<$1\%
(see calibration notes XMM-CCF-REL-189, XMM-CCF-REL-205). 

\begin{table*}
  \caption{Summary of the statistical comparison of EPIC fluxes in different energy bands.} 
\label{table:2}      
\centering                          
\begin{tabular}{c c c c c c c c c c c c c c c c c c }        
\hline\hline                 

Energy band & & pn vs. MOS1 & & & pn vs. MOS2 & & & MOS2 vs. MOS1 \\
 (keV)  & $\mu$(\%) & $\sigma$ & ${\rm N_{srcs}}$ & $\mu$(\%) & $\sigma$ & ${\rm N_{srcs}}$ & $\mu$(\%) & $\sigma$ & ${\rm N_{srcs}}$ \\
\hline                        
      0.2-0.5   & $   -2.0\pm  0.5$ & $ 16.0\pm  0.5 $ &   909  & $  1.2\pm  0.4$ & $ 13.2\pm  0.4 $ &   885  & $ -0.9\pm  0.3$ & $  9.0\pm  0.4 $ &  1092   \vspace{0.05cm}\\
        0.5-1.0   & $   -7.9\pm  0.1$ & $  8.2\pm  0.2 $ &  2132  & $ -8.2\pm  0.1$ & $  7.9\pm  0.2 $ &  2197  & $  1.0\pm  0.2$ & $  7.8\pm  0.3 $ &  2626   \vspace{0.05cm}\\
          1.0-2.0   & $   -8.5\pm  0.2$ & $  7.5\pm  0.2 $ &  2721  & $ -9.1\pm  0.2$ & $  7.3\pm  0.2 $ &  2781  & $  0.3\pm  0.2$ & $  7.2\pm  0.2 $ &  3282   \vspace{0.05cm}\\
        2.0-4.5   & $   -7.1\pm  0.2$ & $  7.1\pm  0.2 $ &  1527  & $ -6.5\pm  0.2$ & $  7.4\pm  0.2 $ &  1546  & $ -0.7\pm  0.2$ & $  7.1\pm  0.2 $ &  1769   \vspace{0.05cm}\\
       4.5-12.0   & $  -12.5\pm  0.5$ & $  9.8\pm  0.5 $ &   484  & $ -9.0\pm  0.5$ & $  9.5\pm  0.5 $ &   502  & $ -3.8\pm  0.2$ & $  8.0\pm  0.2 $ &   549   \vspace{0.05cm}\\
\hline                                   
\end{tabular}

$\mu$ and $\sigma$ are the $\chi^2$ best-fit mean and dispersion of the Gaussian distribution used to fit the data.
${\rm N_{srcs}}$ is the total number of sources involved in the analysis. Errors are 90\% confidence.

\end{table*}

For the MOS cameras new quantum efficiency files (see XMM-CCF-REL-235) 
have become available since the release of {\tt 2XMM}. In addition, a 
significant change in the MOS low energy redistribution characteristics 
with time has been confirmed. This effect is important only for sources 
with a separation from the optical axis $\lesssim$2$\arcm$, which in most 
cases is the target of the observation (Read et al.~\cite{Read06}). To 
account for this effect, epoch-dependent MOS redistribution matrices have 
been generated for use by observers (XMM-CCF-REL-202). However for this 
work we have followed the approach of {\tt 2XMM} and used an average response to compute the MOS 
energy conversion factors. In order to investigate the EPIC flux cross-calibration using the latest calibration available 
we computed new MOS redistribution matrices more representative of the current calibration than the ones used in {\tt 2XMM}.
The MOS1 and MOS2 response matrices were computed with the {\tt SAS} task {\tt rmfgen} for 
revolution 375, on-axis, and for PATTERNs from 0-12. These changes in the MOS calibration have resulted in an increase of 
MOS fluxes compared to the quoted values in {\tt 2XMM} 
by $\sim$5-8\% in the 0.2-0.5 keV band, $\sim$4\% in the 0.5-1.0 keV energy band and $\sim$1\% from 1.0-2.0 keV.
At higher energies the changes in MOS flux are negligible. 
An investigation of the EPIC flux cross-calibration using MOS epoch and position dependent 
responses is beyond the scope of this paper. However as we will see in Sec.~\ref{discussion} 
the use of an 'average' MOS response for all sources has a small effect on the main conclusions of this 
work as the time dependent effect only affects a small fraction of the objects involved in our analysis and 
at low energies only (below $\sim$0.5 keV). 

On-axis MOS and pn effective area files were produced by the {\tt SAS} task {\tt arfgen}. The count rates 
from {\tt emldetect} are corrected for the effective exposure along the FOV (which includes vignetting and bad pixel 
corrections) and the PSF enclosed energy fraction. Hence the effective areas were generated by disabling 
these corrections, as indicated in the documentation of the {\tt SAS} task {\tt arfgen}. 
MOS and pn effective area curves for the medium filter are shown in Fig.~\ref{epic_arf}. For pn we show the effective area 
curves used to compute {\tt ecf} values both above and below 0.5 keV. The energy conversion factors are listed in Table~\ref{table:1}. 

\section{Results}
\label{results}
In order to quantify the systematic difference in flux between the 
EPIC cameras we have performed a statistical analysis, comparing the flux ratio for large samples of sources selected as specified in Sec.~\ref{data}. We computed for each source the value $(S_i-S_j)/S_j$ 
where $S_i$ and $S_j$ are the fluxes of the source measured by cameras ($i$,$j$) respectively.
 
The obtained distributions were fitted with a Gaussian profile, which, thanks to the large number of 
objects involved in the analysis, provided in all cases a good enough representation of the distribution of values. 

\subsection{Energy dependence}
\label{flux_vs_energy}
Fig.~\ref{dists_pn_m1_all} and Fig.~\ref{dists_m2_m1_all} show the 
distributions of difference in flux (in percentage units) when comparing pn and MOS cameras for sources detected 
in different energy bands.
The results of the comparison between pn and MOS2 are very similar to 
those obtained for the pn and MOS1 cameras therefore we only show the latter. The smooth solid lines show the $\chi^2$ best-fit Gaussian 
distributions of the data. The best-fit parameters and number of sources involved in the analysis are shown in the upper-left corner 
of the plots and are listed in Table~\ref{table:2}. Fig.~\ref{disp_all_new_cal} shows the systematic 
difference in the measured flux between cameras as a function of energy.

The mean flux difference between the MOS and pn cameras is less than 10\% at energies $\le$4.5 keV 
and $\sim$10-13\% above 4.5 keV. In all but the very lowest energy cases MOS fluxes are found to be higher than 
those reported for the pn camera. The relative flux calibration between the two MOS cameras is better than 4\% at all energies 
sampled by our analysis. 

In order to obtain the conversion factors from count rate to flux we assumed the same spectral model for all sources. However we know that 
the sources that populate the X-ray sky have a broad range of spectral shapes. For example, the X-ray emission of AGN often shows more 
spectral complexity than a simple power law (excess X-ray absorption, soft excess). Indeed even the distribution of 
broad band continuum shapes is known to have an intrinsic dispersion $\Delta\Gamma$$\sim$0.2-0.3 (see e.g. Mateos et al.~\cite{Mateos05}). 
Furthermore, at soft X-ray 
energies and bright fluxes the contribution from non-AGN populations to the X-ray sky, mainly stars and clusters of galaxies 
with thermal spectra, is not negligible (Mateos et al.~\cite{Mateos08}). 
We have investigated the impact of the variety of spectral shapes on the 
flux cross-calibration of the EPIC cameras. 
In order to quantify the impact of the chosen spectral model on the measured EPIC flux cross-calibration 
we have computed the change in the flux ratio between cameras for a varying power-law spectral model. 
We performed the calculation by varying the continuum shape up to $\Delta\Gamma$=0.6. 
Our simulations have shown that the chosen spectral model has a small effect on the measured flux 
from 0.5 keV to 4.5 keV where a $\Delta\Gamma$=0.6 variation changes the measured flux by less than 4\%.
The EPIC MOS and pn fluxes have a much stronger dependence on the spectral model 
used to compute the {\tt ecf} in the 0.2-0.5 keV and 4.5-12.0 keV energy bands where a $\Delta\Gamma$=0.6 can 
vary the measured fluxes by $\sim$15\%. While the spectral model has an important effect on the measured flux, 
the impact on the flux ratio between cameras is small. 
In order to confirm this result we have also computed the flux ratio distributions 
from the sub-set of sources whose hardness ratios are consistent with the spectral model assumed to
calculate the {\tt ecf}. We found that the results were largely indistinguishable from the `full set' of sources, thus confirming that our approach of using a fixed spectral model to compute source fluxes has a small impact on the main 
results of our analysis.

\begin{table}
  \caption{Grouping definition used to study the time dependence of the EPIC flux cross-calibration.}
\label{table:4}      
\centering                          
\begin{tabular}{c c c c c c c c c c c c c c c c c c }        
\hline\hline                 
Revolution$^a$ &  ${\rm N_{obs}^b}$ \\
\hline                        
 28   -290    & 730  \\
 291  -552    & 919  \\
 553  -814    & 901  \\
 815  -1076   & 739  \\
 1077 -1338   & 202  \\
\hline                                   
\end{tabular}

$^a$ Range of XMM-{\it Newton} revolutions in group.
$^b$ Total number of observations with revolution in group.
\end{table}

\subsection{Time dependence}
\label{flux_vs_time}
Here we present the dependence of the flux cross-calibration of the EPIC cameras on the lifetime of the mission. To do that we have computed the distributions of EPIC 
flux ratios for sources detected at different epochs. We have grouped our sources in five XMM-{\it Newton} 
revolution intervals of the same size to have a uniform sampling of the lifetime of the mission. The grouping definition is 
shown in Table~\ref{table:4} while the observed time dependence of the flux cross-calibration of the EPIC cameras 
at different energies is shown in Fig.~\ref{var_flux_vs_rev}.

The flux cross-calibration between the MOS cameras does not depend significantly on time and the result 
seems to hold at all energies. When comparing pn and MOS fluxes, no significant change of the flux ratio with time is seen 
at energies above 0.5 keV. However, a strong temporal dependence of the flux ratio is found in the 0.2-0.5 keV band, in the 
sense that pn fluxes seem to increase with respect to MOS fluxes at higher revolutions. 
The effect is more important when comparing pn and MOS2 fluxes.
We note that there is some scatter in the pn vs. MOS1 values obtained at different epochs 
at energies above 4.5 keV due to low statistics, although the points are consistent at 2-3$\sigma$.

   \begin{figure}[!t]
   \centering
   \hbox{
   \includegraphics[angle=90,width=0.5\textwidth]{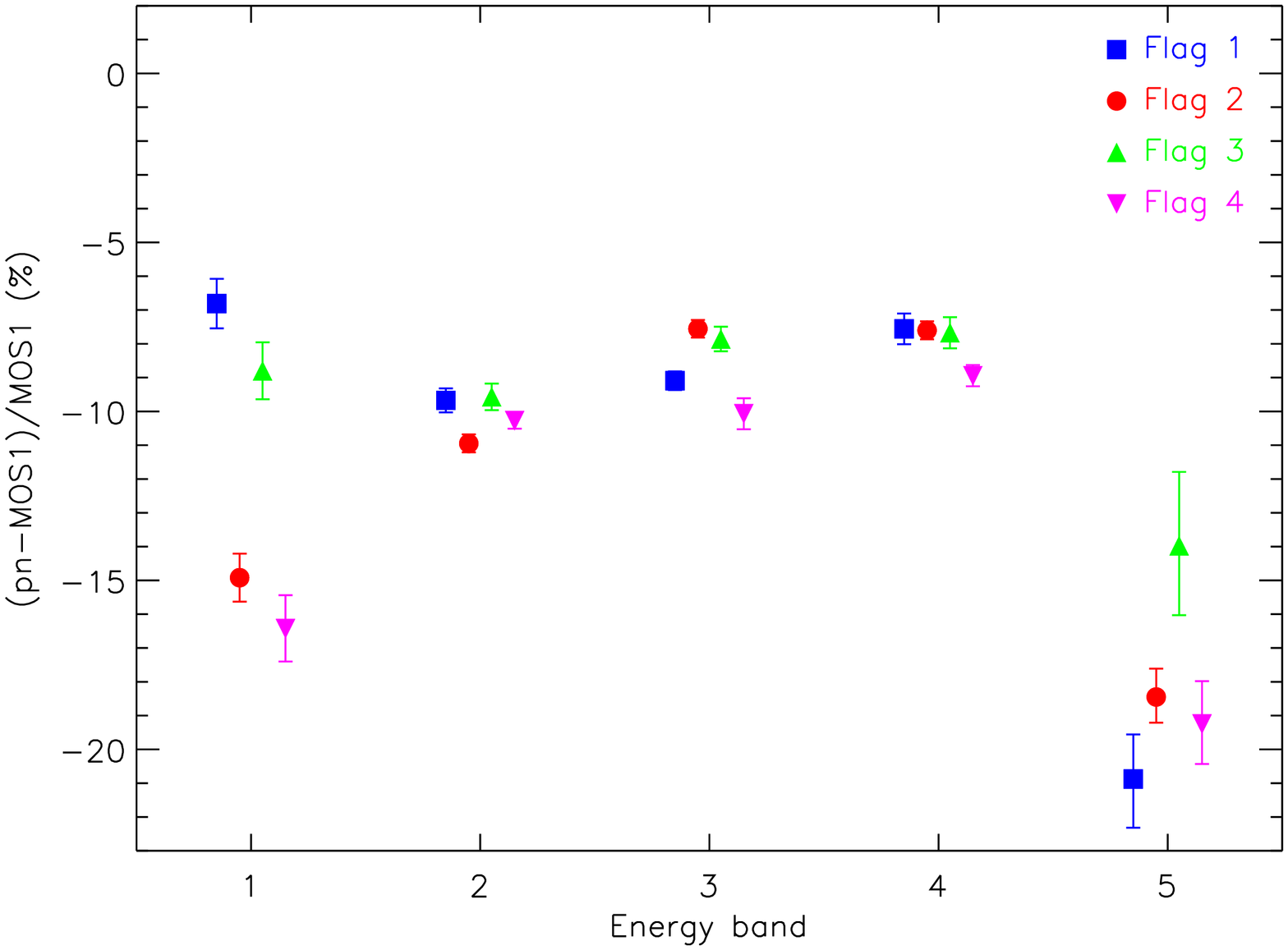}}
   \hbox{
   \includegraphics[angle=90,width=0.5\textwidth]{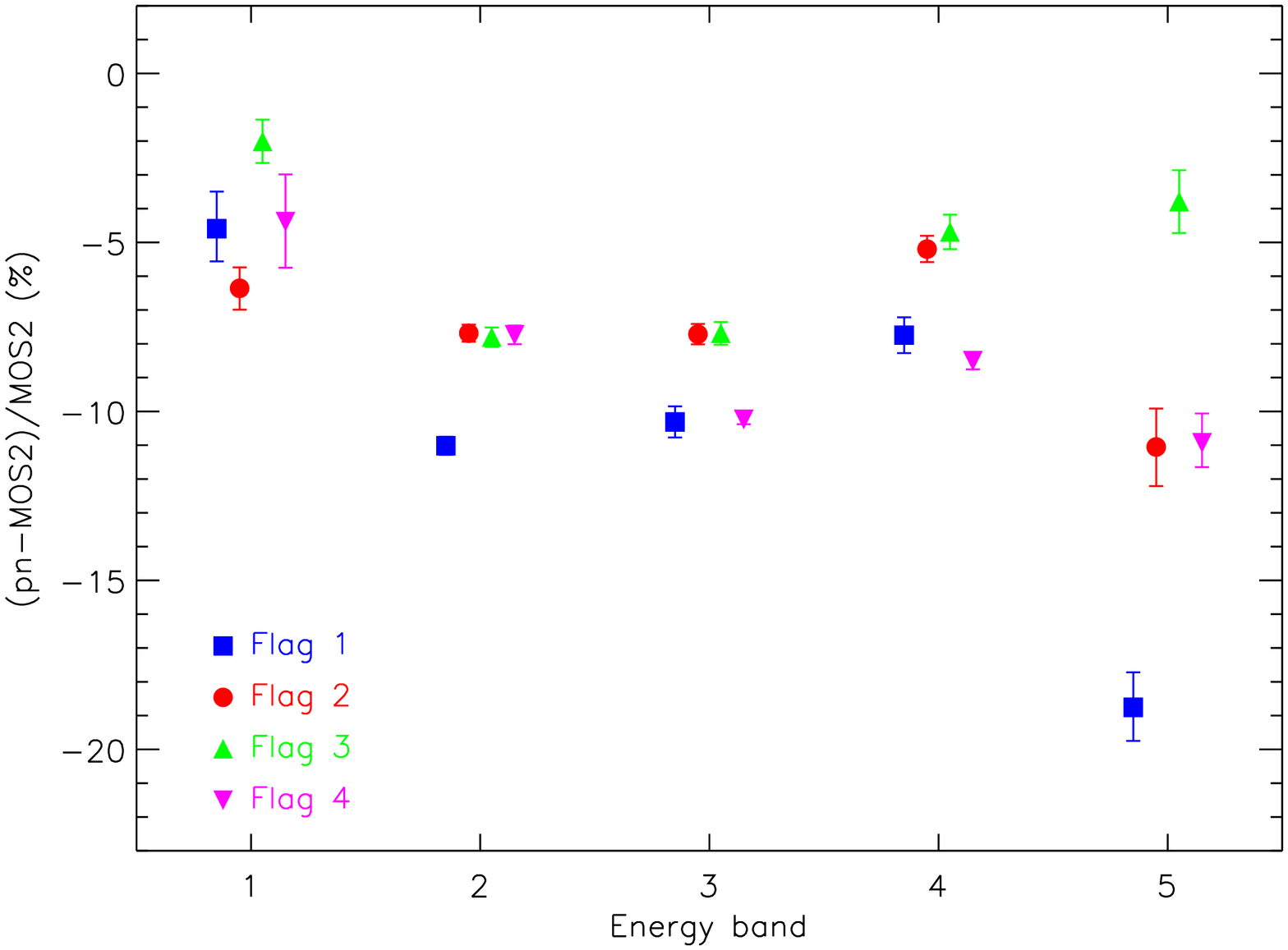}}
   \caption{Variation of the percentage flux difference between the EPIC cameras as a function of the azimuthal position 
     of the sources in the FOV of the MOS cameras (see Sec.~\ref{flux_vs_poss} and Fig.~\ref{mos_flag_def}). Only sources 
     with MOS off-axis 
     values $>$2$\arcm$ have been included 
     in the analysis. 
     The x-axis shows the energy band identification number used in this 
     work: 1: 0.2-0.5 keV, 2: 0.5-1.0 keV, 3: 1.0-2.0 keV, 4: 2.0-4.5 keV and 5: 4.5-12.0 keV. 
     Data points at each energy band have been shifted slightly along the x-axis for clarity.
     Errors are 90\% confidence.}
              \label{var_flux_vs_flag}%
    \end{figure}

   \begin{figure*}[!ht]
   \centering
   \hbox{
   \includegraphics[angle=-90,width=0.48\textwidth]{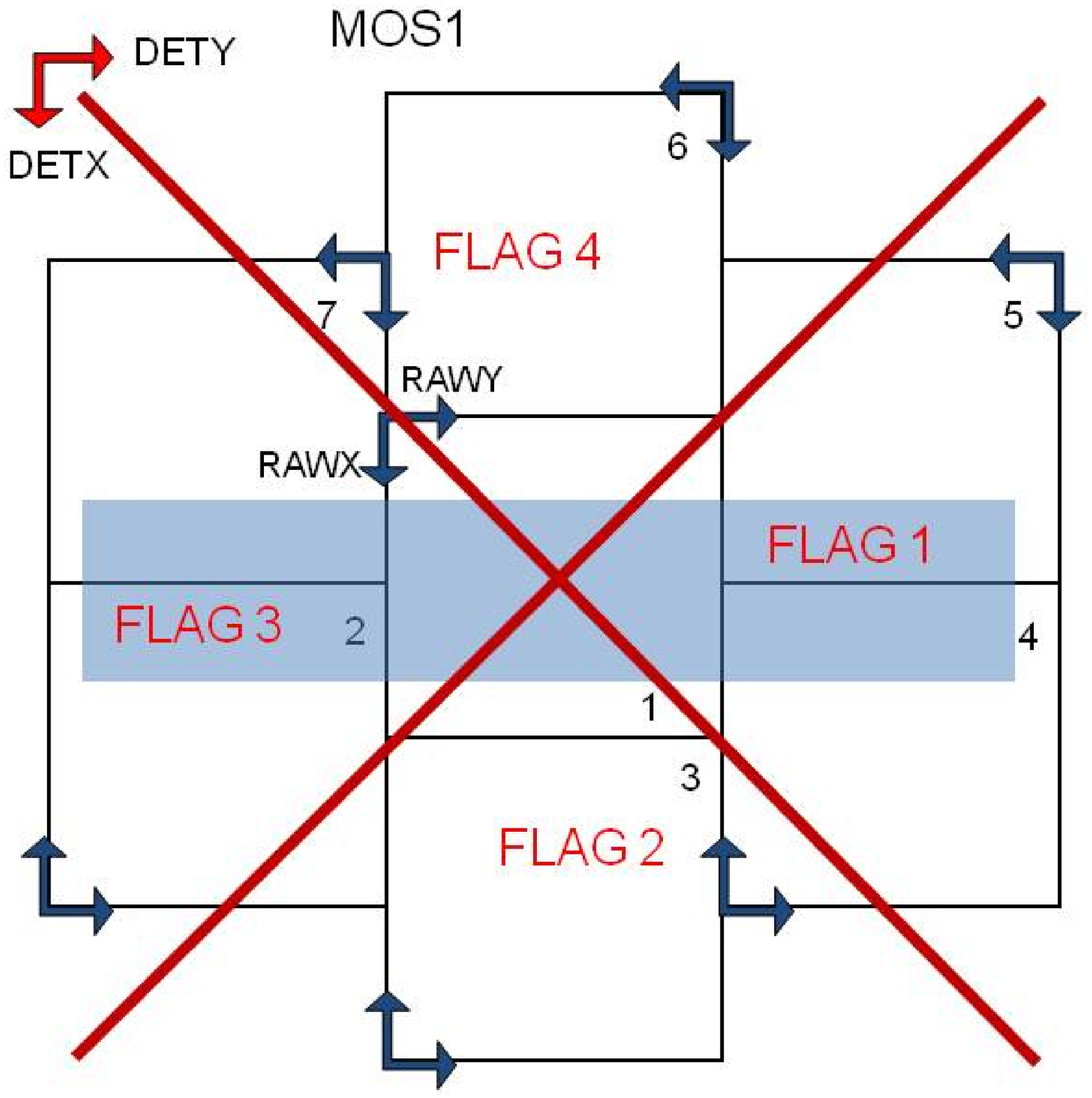}
   \includegraphics[angle=-90,width=0.48\textwidth]{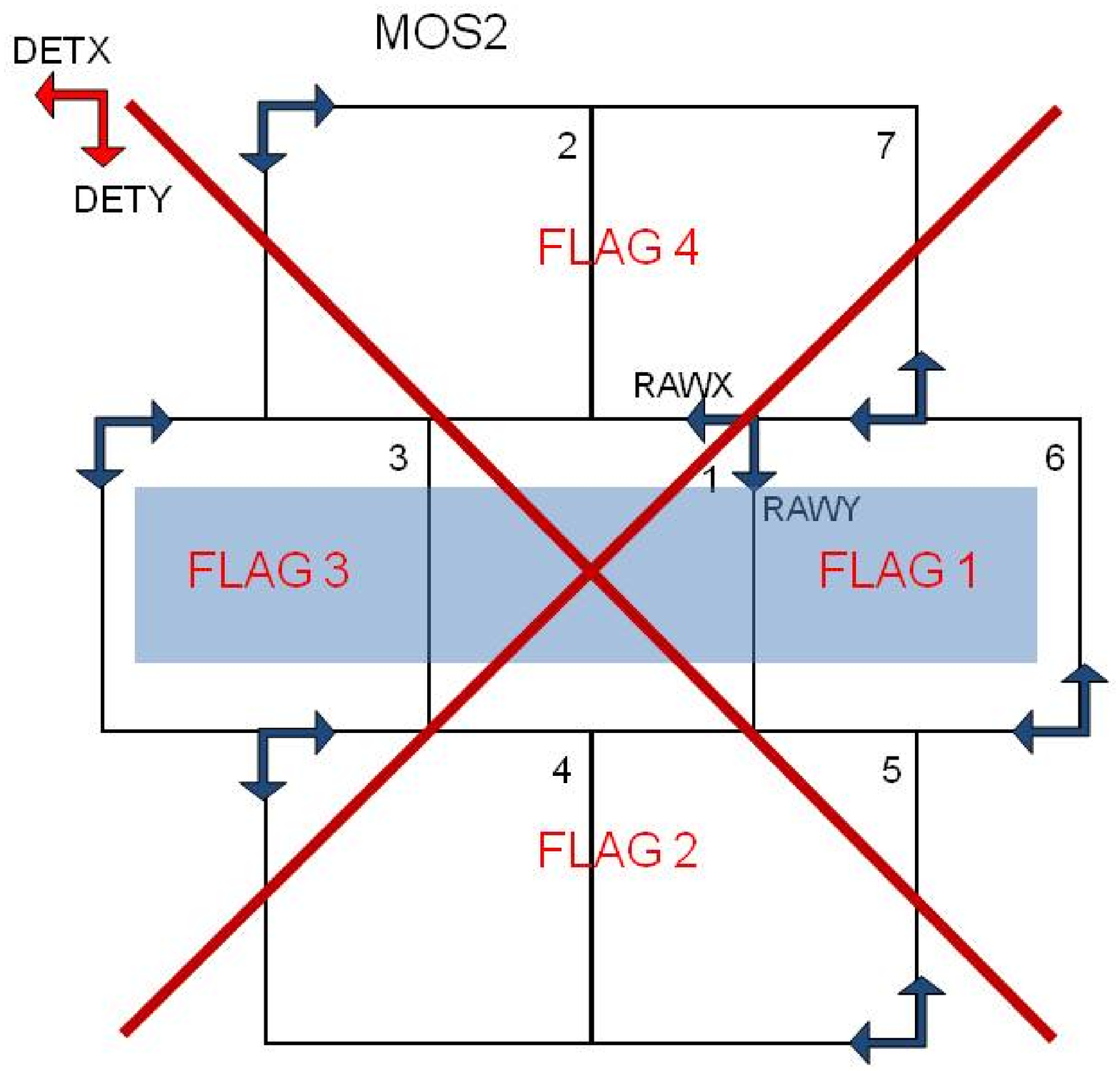}}
   \caption{MOS1 and MOS2 CCD layouts and CCD numbering
(small numbers on the top right corner of the CCDs). The position of the RGSs is indicated with a shaded rectangle. The diagonal cross and large 'FLAG n' 
text indicate the positional flag settings. The MOS diagrams are co-aligned such that the individual sources occupy the same position in each diagram -e.g. a source in the middle of MOS1 CCD 6, lies in the middle of the CCD2/CCD7 boundary in MOS2.}
              \label{mos_flag_def}%
    \end{figure*}

\subsection{Dependence on position of sources in the FOV}
\label{flux_vs_poss}
It is well known that the response of the XMM-{\it Newton} mirrors 
and the EPIC instruments varies with position in the FOV. Here we
investigate the relative flux cross-calibration of the EPIC cameras,
by comparing the EPIC fluxes for sources at different off-axis and
azimuthal angles.

The variation of the percentage flux difference between the EPIC pn
and MOS cameras as a function of energy band for sources at different
MOS off-axis angles is shown in Fig.~\ref{var_flux_vs_offaxis}. Here sources have been divided
into groups with MOS off-axis angles of 0$\arcm$-2$\arcm$, 2$\arcm$-5$\arcm$,
5$\arcm$-8$\arcm$ and 8$\arcm$-12$\arcm$.

There is a clear trend for sources further from the optical axis
to show an increased flux in the MOS cameras relative to pn. This
is particularly apparent for MOS1. In the lowest energy band the on-axis
sources show a significant excess of pn flux relative to the MOS. This 
is a very different result than obtained for sources at larger off-axis angles.
In both MOS cameras, in the highest
energy band the MOS flux excess increases by 10-15\% as the off-axis
angle increases. This highly significant result is likely to be due
to inaccuracies in the calibration of either the vignetting, the PSF or the RGA obscuration.
As a further diagnostic we have produced samples of sources with a MOS
off-axis angle $>$2$\arcm$ (to reduce the effect due to the 
change in the low energy redistribution characteristics of the MOS 
cameras with time, see Sec.~\ref{ecf}) and then divided into four different azimuthal
angle bins. The results are shown in Fig.~\ref{var_flux_vs_flag}.
For both MOS cameras, the positional flags indicate the quadrants
delineated by a diagonal cross centred on CCD=1, RAWX=300, RAWY=300,
such that flag=1 extends towards the low RGA obscuration (i.e. high
MOS throughput) direction, and flag=3 extends towards the high RGA
obscuration (i.e. low MOS throughput) direction, as shown in
Fig.~\ref{mos_flag_def}.

In the 4.5-12.0 keV band there is a very strong dependence of the MOS 
to pn flux ratio on the  
azimuthal angle. For both MOS cameras, sources lying along the RGA
dispersion axis show a large gradient in MOS vs. pn relative flux.

\section{Discussion}
\label{discussion}
The results for the full source sample show an excellent agreement between the two MOS 
cameras; better than 4\% over the entire energy range of the cameras. The MOS cameras 
present a consistent excess in flux compared with the pn of $\sim$7-9\% from 0.5 keV to 4.5 keV and 
$\sim$10-13\% at higher energies ($\gtrsim$4.5 keV). These results are in agreement with the findings of
Stuhlinger et al.~(\cite{Stuhlinger08}), who performed a careful analysis of 168 bright on-axis sources.

At low energies (0.2-0.5 keV) the agreement between the EPIC cameras is seemingly
better ($<$3\%). However, a strong trend in the flux ratios with observation epoch, in this band, is 
evident as shown in Fig.~\ref{var_flux_vs_rev}. This can be explained by the evolution of the MOS redistribution characteristics with time. In this analysis
we have used a single response matrix for each of the MOS cameras, relevant
for an on-axis source
observed in revolution 375, which will necessarily be inaccurate for
observations made at earlier
and later epochs. The redistribution function evolves such
that an increasing fraction of X-rays suffer incomplete charge collection.
The effect is increasingly stronger towards lower energies. The evolution of
the MOS {\tt rmf} seems to be related to the total X-ray radiation 
dose received by a pixel which will clearly be higher at the centre of the
detector where bright sources are preferentially placed (Read et al.~\cite{Read06}). Sources at MOS off-axis angles greater than $\sim$2$\arcm$ are essentially immune to this effect.

For MOS on-axis sources, photons from the energy band 0.2-0.5 keV can be redistributed in
energy below the detection threshold and lost while photons from energy band 0.5-1.0 keV
can be redistributed into the 0.2-0.5 keV energy band. The effect is negligible at energies above $\sim$1 keV. For a given source, the strongest 
evolution in observed flux occurs in the 0.2-0.5 keV band and is dependent upon the intrinsic spectrum. Highly 
absorbed sources which have a very low 0.2-0.5 keV to 0.5-1.0 keV ratio will show an increase in 0.2-0.5 keV
flux as more photons are gained from the 0.5-1.0 keV band than are lost. 

The majority of sources in this sample, however, have relatively soft spectra, for example
AGN with low absorption, and the evolution of the {\tt rmf} causes a loss of 0.2-0.5 keV flux
as more photons are lost below threshold than gained from the 0.5-1.0 keV energy band.
This is confirmed by Fig.~\ref{var_flux_vs_offaxis} where the 0.2-0.5 keV ratio at small off-axis angles
is significantly different from the other points.
We have checked that the distribution of off-axis angles of the sources 
are very similar at all epochs used in our analysis, therefore 
the time dependence observed in the MOS vs. pn 0.2-0.5 keV flux 
ratios (see Fig.~\ref{var_flux_vs_rev}) cannot be explained as an artificial effect due to the 
increasing fraction of near on-axis sources with time.

   \begin{figure}[!t]
   \centering
   \hspace{-0.3cm}\includegraphics[angle=0,width=0.5\textwidth]{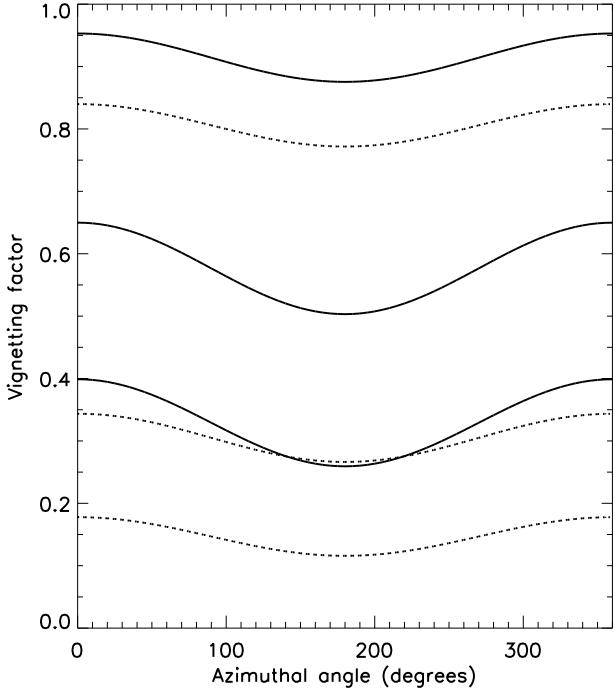}
   \caption{MOS vignetting factor as a function of azimuthal angle at energies of
     1 keV (solid lines) and 10 keV (dotted lines), and for off-axis angles of
     (top to bottom) 3$\arcm$, 9$\arcm$ and 15$\arcm$.
     }
              \label{mos_vignetting}%
    \end{figure}

In Fig.~\ref{var_flux_vs_flag} we see that, once the sources from the
innermost 2$\arcm$ are excluded from the analysis, the 0.2-0.5 keV
flux ratios show a spread of up to $\sim$8\% with the detector
azimuthal angle. As this flux ratio is mostly strongly influenced by
the calibration of the detector {\tt rmf}s we suspect that this spread
is due to intrinsic CCD to CCD variations in the low energy response
of the 14 MOS and 12 pn CCDs. The quality of the ground
calibration data below 500 eV is not sufficient to discriminate
between different devices (XMM-EPIC instrument team, private
communication) so we cannot compare our results directly with a
prediction based on the available calibration data.  Further progress
on this will require the accumulation of sufficient numbers of sources
to enable comparisons to be made on specific combinations of CCDs.

Excluding sources from the innermost 2$\arcm$, the 0.2-0.5 keV flux
ratios are -11.4$\pm$0.3\%, -4.6$\pm$0.6\% and -4.8$\pm$0.7\% for
the pn/MOS1, pn/MOS2 and MOS2/MOS1 ratios respectively bringing them
into better agreement with the values at energies from 0.5 keV to 4.5
keV. It is clear from this that the use of a single MOS {\tt rmf} to
calculate the global flux ratios at the lowest energies is not
correct.

At the highest energies, 4.5-12.0 keV, the MOS1/pn flux excess
increases to 10-13\%. From Fig.~\ref{var_flux_vs_rev} this has no
obvious temporal variation but from Fig.~\ref{var_flux_vs_offaxis} and
Fig.~\ref{var_flux_vs_flag} it is dependent on both the off-axis angle
and the azimuthal angle. The MOS excess increases by $\sim$10\% from
on-axis to off-axis and there is a 10-15\% variation with azimuthal
angle. At these energies the vignetting function and point-spread
function evolve quickly with off-axis angle, suggesting that
inaccuracies in one or both of these calibration quantities may be
responsible. At medium to high energies the ground calibration
data available on individual CCDs suggests that CCD to CCD variations
in detector response are not responsible for the apparent azimuthal
dependence seen in the 4.5-12.0 keV band in Figure 8. For example, the spread in the
measured quantum efficiency of the MOS CCDs is less than $5\%$ from 4.5 to 6.5 keV and
is too small to account for the $15\%$ variation seen in the pn
to MOS2 flux ratio in the 4.5-12.0 keV band. In principle the pn camera has no azimuthal
dependence for any of the calibration quantities, whereas the
vignetting of the MOS cameras is strongly affected by the transmission
of the RGS gratings (Turner et al.~\cite{Turner01}). Furthermore, because the
outer MOS chips lie on two distinct planes, it is believed that small
$< 5$\% variations due to the PSF may exist (Saxton et
al.~\cite{Saxton03}).

In Fig.~\ref{mos_vignetting} we show the current calibration of the MOS vignetting
function for off-axis angles of 3$\arcm$, 9$\arcm$ and 15$\arcm$ as a function of the
azimuthal angle. The azimuthal dependency, due to obscuration by the
RGA structure, is both energy and off-axis angle dependent. At 10 keV and
9$\arcm$ off-axis it drops by 25\% between the high MOS throughput
direction (azimuth=$0^{\circ}$; MOS flag 1 in Fig.~\ref{var_flux_vs_flag} 
and Fig.~\ref{mos_flag_def}) and the opposite
direction (azimuth=$180^{\circ}$; MOS flag 3 in Fig.~\ref{var_flux_vs_flag} and Fig.~\ref{mos_flag_def}). 
Therefore the results presented in Fig.~\ref{var_flux_vs_flag} are highly suggestive that a 
calibration change is required: increasing 
the magnitude of the azimuthal vignetting variation at high energies would bring the points 
in Fig.~\ref{var_flux_vs_flag} closer together.
Furthermore, a change in the radial vignetting calibration to increase the
overall MOS throughput at the highest energies might also be
required. This is especially the case for MOS1 where a large overall
offset at high energies is seen in Fig.~\ref{disp_all_new_cal}.


Though the effect introduced by the outer MOS chips lying on two
distinct planes is thought to be small, it can be examined in the
present analysis. The outer MOS chips run up-down, up-down around the
central chip, such that the positional flag settings used in 
Fig.~\ref{var_flux_vs_flag} and Fig.~\ref{mos_flag_def} divide MOS1 into two 
flags (2 and 4) that are dominated by
an `up' chip and by a `down' chip, and are free from the gross effects of the azimuthal vignetting (at
azimuth=$0^{\circ}$ and azimuth=$180^{\circ}$). The 
situation is not as good for MOS2 as,
although CCDs 6 and 3 can be compared by comparing flags 1 and 3,
these positions suffer the largest deviations in azimuthal vignetting
due to the RGA obscuration. For MOS1 at least though we can examine
the MOS chip plane PSF effect on the obtained source fluxes:
the flux ratio values in these two areas (MOS1, flags 2 and 4) in Fig.~\ref{var_flux_vs_flag} are seen to agree with each other 
very well (within 2\%), indicating
that the effect on the outer MOS chips lying on two distinct planes is very small.

\section{Conclusions}
\label{conclussions}
We have used the second XMM-{\it Newton} serendipitous source catalogue, {\tt 2XMM} to compile 
large samples of `good quality' sources to measure the
difference in flux seen by the EPIC cameras and investigate the
relative calibration of the instruments. The main results are given below.

\begin{enumerate}
\item The MOS cameras agree with each other to better than 4\% at all energies.

\item The MOS cameras register a consistent 7-9\% higher flux than the pn at energies between 0.5 keV and 4.5 keV.

\item The pn/MOS flux discrepancy increases at high energies, up to $\sim$13\% for MOS1. 
  The flux discrepancy at high energies has been shown to have a strong
  off-axis and azimuthal angle dependence. The most likely explanation for this
  variation is an incorrect calibration of the obscuration factor in the
  MOS cameras due to the RGA structures. As the azimuthal variation is not
  present at energies below 4.5 keV, a reworking of the RGA physical parameters is needed 
  which solves the high energy problem without affecting lower energies. If this can be achieved 
  then the mirror vignetting function for the EPIC MOS cameras will need to be revisited as these
  two quantities are intimately related. Ideally, the solution would
  reduce the overall discrepancy of the cameras at higher energies and
  leave a uniform MOS/pn excess at all energies.

\item A systematic difference in flux cross-calibration is seen for the 0.2-0.5 keV band with a strong dependence on time. 
  This effect, also present in the {\tt 2XMM} catalogue, 
  has been explained as being due to the fact that an average response has been used to compute the fluxes 
  of sources detected with the MOS camera at all epochs. Therefore this effect could be removed by 
  using an epoch and position dependent response function to compute the MOS1 and MOS2 fluxes and could be used 
  in future releases of XMM-{\it Newton} catalogues. The fidelity of fluxes in the XMM-{\it Newton} source catalogues could 
  be further improved by using the source spectrum, or hardness ratio, to calculate the {\tt ecf}.
\end{enumerate}

\begin{acknowledgements}
We thank Matteo Guainazzi, Martin Stuhlinger and Simon Rosen for useful comments.
We gratefully acknowledge the XMM-{\it Newton} Survey Science Centre, a
consortium of 10 European institutes, for the production and public
provision of the {\tt 2XMM} catalogue.
We would like to acknowledge the contribution of all the calibrators of the EPIC and RGS
instruments on XMM-{\it Newton}, who have helped to achieve the self-consistency between
the instruments that we see today.
SM, AMR and SS acknowledge direct support from the UK STFC research council.
We thank the referee for providing comments that improved this paper.

\end{acknowledgements}



\begin{thebibliography}{}

   \bibitem[2001]{Herder01} den Herder, J. W., Brinkman, A. C., Kahn, S. M., Branduardi-Raymont, G., et al. 2001 
     A\&A, 365, 7


   \bibitem[2001]{Ehle01} Ehle, M., Altieri, B. 2001, SOC note XMM-SOC-CAL-PL-0001 Erd C. 2000, SOC note XMM-SOC-PS-TN-0038


   \bibitem[2001]{Jansen01} Jansen, F., Lumb, D., Altieri, B., Clavel, J., et al. 2001
     A\&A, 365, 1

   \bibitem[1997]{Kendziorra97} Kendziorra, E., Bihler, E., Grubmiller, W., Kretschmar, B., et al. 1997
     SPIE, 3114, 155

   \bibitem[1999]{Kendziorra99} Kendziorra, E., Colli, M., Kuster, M., Staubert, R., et al. 1999
     SPIE, 3765, 204

   \bibitem[1999]{Kuster99} Kuster, M., Benlloch, S., Kendziorra, E., Briel, U. G. 1999
     SPIE, 3765, 673

   \bibitem[2002]{Lumb02} Lumb, D. H., Warwick, R. S., Page, M. \& De Luca, A. 2002,
     A\&A, 389, 93

   \bibitem[2005]{Mateos05} Mateos, S., Barcons, X., Carrera, F. J., Ceballos, M. T., et al. 2005
     A\&A, 444, 79

   \bibitem[2008]{Mateos08} Mateos, S., Warwick, R. S., Carrera, F. J., Stewart, G. C., et al. 2008
     A\&A, 492, 51

   \bibitem[2006]{Read06} Read, A. M., Sembay, S. F., Abbey, T. F., Turner, M. J. L. 
     Proceedings of the "The X-ray Universe 2005", 26-30 September 2005, El Escorial, Madrid, Spain. Ed. by A. 
     Wilson. ESA SP-604, Volume 2, Noordwijk: ESA Publications Division, ISBN 92-9092-915-4, 2006, p. 925 - 929

   \bibitem[2003]{Saxton03} Saxton, R. D., Denby, M., Griffiths, R. G., Neumann, D. M. 2003
     AN, 324, 138

   \bibitem[2001]{Struder01} Str\"{u}der, L., Briel, U., Dennerl, K., Hartmann, R., et al. 2001
     A\&A, 365, 18

   \bibitem[2008]{Stuhlinger08} Stuhlinger, M., Kirsch, M.G.F., Santos-Leo, M., Pollock, A.M.T., et al. 2008
     `Status of the XMM-Newton instrument cross-calibration with SASv7.1', XMM-SOC-CAL-TN-0052.

   \bibitem[2001]{Turner01} Turner, M. J. L., Abbey, A., Arnaud, M., Balasini, M., et al. 2001,
     A\&A, 365, L27-L35

   \bibitem[2008]{Watson08} Watson, M.G., Schr\"{o}der, A.C., Fyfe, D., Page, C.G. 2008,
     A\&A, 493, 339

\end{thebibliography}
\end{document}